\documentclass[11pt]{article}

\usepackage{cite}
 \usepackage{subfigure}
\usepackage{helvet}
\usepackage{amsmath}
\usepackage{amssymb}
\usepackage{setspace}
\usepackage{graphicx}
\usepackage{empheq}
\usepackage{soul}
\usepackage{tabularx}
\usepackage{array}
\usepackage{float}
\usepackage{braket}
\usepackage[utf8]{inputenc}
\usepackage{url}
\usepackage{hyperref}
\usepackage{slashed}
\usepackage[font=footnotesize,labelfont=bf]{caption}

\def\be{\begin{equation}}
\def\ee{\end{equation}}

\usepackage[a4paper,top=3cm,bottom=3cm,left=2.5cm,right=2.5cm,bindingoffset=0mm]{geometry}

\begin{document}

\begin{center}
{\Large \bf Physics with Leptons and Photons at the LHC}

\vspace*{1cm}
                                                   
L. A. Harland--Lang \\                                                 
                                                   
\vspace*{0.5cm}
Rudolf Peierls Centre, Beecroft Building, Parks Road, Oxford, OX1 3PU                                          
                                                    
\vspace*{1cm}

\begin{abstract}
\noindent
We present a phenomenological study of photon--initiated (PI) lepton production at the LHC, as implemented in the structure function (SF) approach. We provide detailed predictions for multi--differential lepton pair production, and show that the impact on observables sensitive to the the weak mixing angle, $\sin^2 \theta_W$, and $W$ boson mass, $M_W$, as well as PDFs can be non--negligible, in particular given the high precision being aimed for. The SF calculation can provide percent level precision in the corresponding production cross sections, and is therefore well positioned to account for LHC precision requirements. We consider the pure $\gamma\gamma$ channel, and compare in detail to the NLO collinear calculation. We  in addition include initial--state $Z$ as well as mixed $\gamma /Z + q$ contributions, and assess their impact. We also consider photon--initiated lepton--lepton scattering, and again find the SF approach can provide high precision predictions for this process in way that can straightforwardly account for any fiducial cuts imposed. Finally, we provide a publicly available Monte Carlo generator, \texttt{SFGen}, for PI lepton pair production and lepton--lepton scattering within the SF approach, for use by the community.
  
\end{abstract}

\end{center}

\section{Introduction}

A major aim of the LHC, and the high--luminosity upgrade (HL--LHC) that will follow, is to precisely test the Standard Model (SM) predictions for as wide a range of collider processes as possible. A particularly important element of this involves events with leptons in the final state. To give two key examples, the Drell Yan production of a lepton pair measured multi--differentially allows a measurement of the weak mixing angle, $\sin^2 \theta_W$, as well as providing constraints on the parton distribution functions (PDFs) of the proton, while  precision data on $W$ production via leptonic decays can provide a correspondingly precise measurement of the $W$ mass, $M_W$. Significant amounts of such data have already been collected at the LHC, see e.g.~\cite{Aaij:2015gna,Aaij:2015zlq,Aaij:2015vua,Khachatryan:2016pev,Aaboud:2016btc,Aaboud:2017svj,Aaboud:2017ffb,Sirunyan:2018swq,Aad:2019rou,Sirunyan:2019bzr}. In more detail, the ATLAS 7 TeV analysis~\cite{Aaboud:2017svj} presents a measurement of $M_W$ with a total uncertainty of $19$ MeV, which is comparable to previous Tevatron determinations. High precision triple--differential measurements of lepton pair production have also recently been reported at 8 TeV by both ATLAS~\cite{Aaboud:2017ffb} and CMS~\cite{Sirunyan:2018swq}, with the latter reporting a value of $\sin^2 \theta_W$ that is beginning to bear down on LEP precision. As discussed in~\cite{Azzi:2019yne}, significant improvements on the precision of these measurements are anticipated and aimed for in the future, in particular from HL--LHC running.

A key ingredient in this programme is the availability of high precision theoretical predictions for the SM production processes. An important and topical element of this is the contribution from photon--initiated (PI) channels to lepton pair production. A great deal of theoretical progress has occurred in the modelling of this process, until recently within the context of including the photon as an additional partonic constituent of the proton, that is via a photon PDF.  Following earlier work of~\cite{Martin:2004dh,Schmidt:2015zda,Ball:2013hta,Giuli:2017oii,Harland-Lang:2016apc,Harland-Lang:2016kog}, the LUXqed group~\cite{Manohar:2016nzj,Manohar:2017eqh} applied the method originally appearing in the context of the equivalent photon approximation (EPA)~\cite{Budnev:1974de}  for determining the photon flux from the elastic and inelastic proton structure functions, providing the first publicly available PDF set within this approach (see also~\cite{Anlauf:1991wr,Blumlein:1993ef,Mukherjee:2003yh,Luszczak:2015aoa} for earlier work). Following from this, photon PDFs have been provided in combination with the global MMHT~\cite{Harland-Lang:2019pla} and NNPDF~\cite{Bertone:2017bme} sets, applying the same approach as LUXqed. In all cases, the experimental input for the corresponding structure function inputs is sufficiently precise that the quoted photon PDF uncertainty is very small, generally at the $\sim 1 \%$ level, and thus a high precision photon PDF determination can be claimed.

This however is not the end of the story. While the photon PDF itself is known with high precision, the corresponding PI cross section has at leading order (LO)  a sizeable factorization scale variation uncertainty associated with it, that is significantly larger than the small PDF uncertainty. While this situation can be, and should be, improved upon by working to NLO or beyond, as discussed in detail in~\cite{Harland-Lang:2019eai}, an alternative approach is to work in the `structure function' (SF) framework. Here, one calculates the cross section directly in terms of the proton structure functions, in an analagous manner to the calculation of Higgs boson production via vector boson fusion (VBF)~\cite{Han:1992hr,Figy:2003nv,Bolzoni:2010xr,Cacciari:2015jma,Dreyer:2016oyx,Cruz-Martinez:2018rod}.
This contains no reference to the photon PDF, and it was demonstrated in~\cite{Harland-Lang:2019eai} for the specific cases of a toy lepton--hadron scattering processes and off--$Z$ peak lepton pair production at the LHC that this provides predictions which are by construction more precise than calculations within the standard collinear factorization approach. Percent level precision in the predicted cross sections was in particular achieved for the first time, with remaining contributions from e.g. the `non--factorizable' corrections that break the structure function picture expected to be small, as they are in the VBF Higgs case (see~\cite{Liu:2019tuy,Dreyer:2020urf}). 

In this work, we build on this previous study in a number of directions: we investigate in more detail the implications for precision phenomenology at the LHC, calculating the PI contribution to multi--differential dilepton production, which is relevant for determinations of $\sin^2\theta_W$ as well as PDF constraints, and evaluating the potential impact  on $W$ mass determinations, through the tuning that is performed to the $Z$ event sample, from which the PI contribution must be subtracted; we include contributions from $Z$ bosons in the initial state, as well as photons, and the mixed $Z/\gamma +q$ contributions, within the SF framework, which while generally  small can be important in some regions; we calculate the PI cross section for lepton--lepton scattering and consider the interplay this has with the recently developed lepton PDF formalism~\cite{Buonocore:2020nai}; finally, we provide the above results in the publicly available \texttt{SFGen} Monte--Carlo (MC) generator, for use by the community.

In more detail, we first investigate the relative impact of PI on the dilepton rapidity distribution at the LHC, which enters at the level of $5-10\%$ of the QCD DY contribution away from the $Z$ peak region. We in addition compare in detail the high precision SF results for the pure $\gamma\gamma$ PI process, which dominates away from the $Z$ peak, with the calculation in collinear factorisation, at both LO and NLO; the latter order corresponds to the highest currently available precision calculation within the collinear approach. The LO collinear predictions as expected exhibit large scale variation uncertainties, which are significantly larger than the small PDF uncertainty on the photon PDF, and the related experimental uncertainty on the SF calculation, due to the determination of the structure functions. If we take $\mu=m_{ll}$ for the renormalization/factorization scale, the agreement between the NLO collinear prediction and the SF is indeed improved with respect to the LO case, and the scale variation uncertainties reduced, although at low to intermediate masses these remain significantly larger than the PDF uncertainty. Moreover, taking a differing scale choice related to the lepton $p_\perp^l$, the convergence of the NLO collinear result is greatly reduced, and the agreement with the full SF result rather poor. Thus, we find that while in general the inclusion of NLO corrections can improve the precision and accuracy of the collinear result, the overall picture is not completely clear cut; in some cases there remain reasonably large scale variation uncertainties and areas of disagreement with the full SF result, depending on the choice of scale and/or experimental cuts.

We will in addition show that the PI contribution of the off--$Z$ peak ATLAS 8 TeV triple differential lepton pair data~\cite{Aaboud:2017ffb} is expected to enter at up to the $\sim 6\%$ level in comparison to the QCD DY cross section, with a non--trivial impact on the corresponding $\cos \theta^*$ distribution. Similar results will be expected for other measurements of this process. This can therefore potentially impact on $\sin^2 \theta_W$ determinations as well as PDF constraints. 

We also consider the impact of PI contributions on dilepton production in the $Z$ peak region, showing that while inclusively these enter at the per mille level, at lowest $p_\perp^{ll}$ values this is rather larger, at the $\sim 0.5\%$ level. This is a result of the suppression of QCD DY process in this region, and the enhancement of PI production, due in part to the contribution from elastic photon emission; this can only be predicted reliably by using the SF approach directly. Given the level of precision being aimed for in $W$ mass determinations and the fact that the dilepton low $p_\perp^{ll}$ region is tuned to in order to extract $M_W$ from  $l\nu$ events (for which PI production of this sort is absent), this may in principle have an impact on such determinations. That is, the irreducible PI component must be subtracted from the dilepton tuning sample,  and the effect of this necessarily depends on the modelling of PI production.  We provide a quantitative estimate of this possible impact, and show that while small it is not necessarily negligible with respect to the uncertainties in $M_W$ being aimed for at the LHC.

The method for including contributions from $Z$ bosons in the initial state, and the mixed $Z/\gamma +q$ contributions follows from a straightforward generalisation of the pure $\gamma\gamma$ case. As mentioned above, the impact from these is generally small, but this is not alway true: for example, at higher dilepton invariant masses the contribution from initial--state $Z$ bosons is not negligible, while when considering the lepton pair $p_\perp^{ll}$ distribution,  $Z/\gamma +q$ contributions can be important. The treatment of these contributions raises a wider point, namely that the benefit of applying the SF approach directly, while transparent for process where the final state of interest ($l^+ l^-$, $W^+ W^-$...) is directly produced by the $\gamma\gamma$ initial state, is less clear in the mixed $\gamma + q$ case. Here, one must deal with the collinear enhancement of the $\gamma \to q\overline{q}$ splitting, and at this level of precision include QED DGLAP evolution of the quark/antiquark PDFs, which certainly requires one introduce a photon PDF within the LUXqed approach.  Nonetheless, one may systematically account for this while still working within the SF approach. We in particular discuss the complication that arises due to double counting with the PI contribution that is already effectively included via the QED DGLAP evolution of the quark PDFs, and provide a simple procedure for removing this. The formal precision of the SF calculation is however then tied to the order at which one does this subtraction.

In a recent study~\cite{Buonocore:2020nai} it has been demonstrated how a lepton originating from a $\gamma^* \to l^+ l^-$ splitting can be viewed as an initial--state parton in the production process, and hence accounted for via a lepton PDF in the proton, which is determined to high precision in a manner analogous to the LUXqed case for the photon (see~\cite{Bertone:2015lqa} for earlier work). As with the photon PDF, one can instead calculate the corresponding cross section in the SF approach directly, including the $\gamma^* \to l^+ l^-$ transition via this. We will demonstrate this for the specific example of same--sign same--flavour, or different--flavours arbitrary--sign lepton pair production. We find that the SF approach can indeed provide high precision predictions for this process, and the result of applying cuts in order to isolate back--to--back lepton topologies is readily evaluated. 

Finally, the \texttt{SFGen} MC provides a publicly available tool for lepton pair and lepton--lepton scattering within the SF approach, including initial--state $Z$ and mixed $Z/\gamma +q$ contributions. Arbitrary distributions and unweighted events can be generated, and errors due to the experimental uncertainty on the structure function (the equivalent of PDF uncertainties in the photon PDF framework) can be calculated on--the--fly. Unweighted events an be interfaced to \texttt{Pythia} for showering/hadronization of the proton dissociation system, following the approach discussed in~\cite{Harland-Lang:2020veo}. 

The outline of the paper is as follows. In Section~\ref{sec:gamzq} we overview the SF approach and discuss the inclusion of initial--state $Z$ contributions. In Section~\ref{sec:zq}
we show how  mixed $Z/\gamma +q$ contributions are included. In Section~\ref{sec:invm} we present results for the dilepton invariant mass distributions.  In Section~\ref{sec:nlo} we present a comparison for these distributions between the SF calculation and the collinear at LO and NLO.  In Section~\ref{sec:dileppt} we consider the impact on the dilepton $p_\perp^{ll}$ distribution. In Section~\ref{sec:mw} we consider the potential impact of the PI contribution to dilepton production on $W$ boson mass determinations. In Section~\ref{sec:ll3d} we present results for the PI contribution to triple--differential lepton pair production. In Section~\ref{sec:llscat} we provide predictions for lepton--lepton scattering and compare with the lepton PDF framework. In Section~\ref{sec:SFgen} we describe the \texttt{SFGen} MC and its availability. Finally, in Section~\ref{sec:conc} we conclude.

\section{Theory}

\subsection{Photon and $Z$ boson initiated production}\label{sec:gamzq}

The  expression for the photon--initiated cross section in proton--proton collisions is given in~\cite{Harland-Lang:2019eai}, which itself is based on the analysis of~\cite{Budnev:1974de}. Here we straightforwardly generalise this to include $Z$ exchange and $Z/\gamma$ interference:
  \be\label{eq:sighh}
  \sigma_{pp} = \sum_{Y_1,Y_2}\frac{1}{2s}  \int \frac{{\rm d}^3 p_1' {\rm d}^3 p_2' {\rm d}\Gamma}{E_1' E_2'}  \alpha(Q_1^2)\alpha(Q_2^2)
  \frac{\rho_{Y_1,1}^{\mu\mu'}\rho_{Y_2,2}^{\nu\nu'} \left[M^*_{\mu'\nu'}M_{\mu\nu}\right]_{Y_1,Y_2}}{Q_1^2 Q_2^2}\delta^{(4)}(q_1+q_2 - k)\;,
 \ee
 where $Y_i= \gamma\gamma, Z\gamma, ZZ$ corresponds to pure PI, $\gamma Z$ interference and pure $Z$--initiated production, respectively. The outgoing hadronic systems have momenta $p_{1,2}'$ and the photons have momenta $q_{1,2}$, with $q_{1,2}^2 = -Q_{1,2}^2$. We consider the production of a system of 4--momentum $k = q_1 + q_2 = \sum_{j=1}^N k_j$ of $N$ particles, where ${\rm d}\Gamma = \prod_{j=1}^N {\rm d}^3 k_j / 2 E_j (2\pi)^3\equiv  \prod_{j=1}^N {\rm d}\Gamma_j$ is the standard phase space volume. $M^{\mu\nu}$ corresponds to the relevant $VV \to X(k)$ production amplitude, with $V=\gamma,Z$ and the `$Y_1, Y_2$' subscript indicating that  pure $Z$, $\gamma$ or $Z/\gamma$ interference be included in the appropriate way. $\rho$ is the density matrix, which is given in terms of the well known proton structure functions:
  \begin{align}\nonumber
 \rho_{Y_i,i}^{\alpha\beta}=2\eta_{Y_i}\int \frac{{\rm d}M_i^2}{Q_i^2} &\bigg[-\left(g^{\alpha\beta}+\frac{q_i^\alpha q_i^\beta}{Q_i^2}\right) F_1^{Y_i}(x_{B,i},Q_i^2)+ \frac{(2p_i^\alpha-\frac{q_i^\alpha}{x_{B,i}})(2p_i^\beta-\frac{q_i^\beta}{x_{B,i}})}{Q_i^2}\frac{ x_{B,i} }{2}F_2^{Y_i}(x_{B,i},Q_i^2)\\ \label{eq:rho}
 &-i\epsilon^{\alpha\beta\mu\nu}q_{i,\mu} p_{i,\nu} \frac{ x_{B,i} }{Q_i^2}F_3^{Y_i}(x_{B,i},Q_i^2)\bigg]\;,
 \end{align}
where $p_i$ is the proton $i$ momentum, $x_{B,i} = Q^2_i/(Q_i^2 + M_{i}^2 - m_p^2)$ for a hadronic system of mass $M_i$ and we note that the definition of the photon momentum $q_i$ as outgoing from the hadronic vertex is opposite to the usual DIS convention. Here, the integral over $M_i^2$ is understood as being performed simultaneously with the phase space integral over $p_{i}'$, i.e. is not fully factorized from it (the energy $E_i'$ in particular depends on $M_i$). The prefactor $\eta$ is given by
\be
\eta_{\gamma\gamma,i} = 1\;,\qquad \eta_{\gamma Z, i } = \frac{G_F M_Z^2}{2 \sqrt{2}\pi \alpha}\frac{Q^2_i}{Q_i^2 + M_Z^2}\;,\qquad \eta_{ZZ, i} = \eta_{\gamma Z, i }^2\;.
\ee
For the structure functions, we include the corresponding elastic and inelastic contributions in the $\gamma\gamma$ case and include an uncertainty due to the experimental inputs on these, as described in~\cite{Harland-Lang:2019eai}. These are evaluated following the procedure discussed in~\cite{Harland-Lang:2019pla}, which is closely based on that described in~\cite{Manohar:2016nzj,Manohar:2017eqh}. We refer the reader to these references for further details, but in summary we include: an uncertainty on the A1 collaboration~\cite{Bernauer:2013tpr} fit to the elastic proton form factors, based on adding in quadrature the experimental uncertainty on the polarized extraction and the difference between the unpolarized and polarized; a $\pm 50\%$ variation on the ratio $R_{L/T}$, relevant to the low $Q^2$ continuum inelastic region; a variation of $W^2_{\rm cut}$, the scale below which we use the CLAS~\cite{Osipenko:2003bu} fit to the resonant region, and above which we use the HERMES~\cite{Airapetian:2011nu} fit/pQCD calculation (for $Q^2$ below/above $1\,{\rm GeV}^2$), between $3$--$4$ ${\rm GeV}^2$; the symmetrised difference between the default CLAS and Cristy--Bosted~\cite{Christy:2007ve} fits to the resonant region; the standard PDF uncertainty on the \texttt{MMHT2015qed\_nnlo} quark and gluon partons in the $Q^2_i > 1 \,{\rm GeV}^2$ continuum region, as calculated via NNLO in QCD  predictions for the structure functions in the ZM--VFNS,  implemented in~\texttt{APFEL}~\cite{Bertone:2013vaa}. We do not include any uncertainty due to non--factorizeable corrections: in the case of VBF Higgs (and Higgs pair) production these are found to enter at the percent level or less~\cite{Liu:2019tuy,Dreyer:2020urf}, and while no calculation is currently available for PI production, we can expect a roughly similar result here.

For $Z\gamma$ and $ZZ$ structure functions, bearing in mind that these will only be at all relevant for higher $Q^2 \sim M_Z^2$, we can safely ignore any elastic or low $Q^2$ resonant contributions. We therefore  simply set the structure functions to zero for $Q^2_i < 1 \,{\rm GeV}^2$, and the uncertainty in this case is purely due to the PDF uncertainty on the pQCD prediction. From the above expressions, one naturally expects that at lower $Q_i^2$ the $Z/\gamma$ contribution will be strongly suppressed by $\sim Q_i^2/M_Z^2$, and hence will generally not be significant. However, as we will see at larger values of $m_{ll}$, where the phase space for larger $Q_i^2$ is increased, and/or in kinematic regions where larger $Q_i^2$ is preferred, such as at larger $p_\perp^{ll}$, they can play an important role.

For all numerical evaluations we as in~\cite{Harland-Lang:2019zur} write \eqref{eq:sighh} as
  \be\label{eq:sighhvar}
  \sigma_{pp} = \sum_{Y_1,Y_2}\frac{1}{2s}   \int  {\rm d}x_1 {\rm d}x_2\,{\rm d}^2 q_{1_\perp}{\rm d}^2 q_{2_\perp} {\rm d}\Gamma \alpha(Q_1^2)\alpha(Q_2^2)\frac{1}{\tilde{\beta}}
  \frac{\rho_{Y_1,1}^{\mu\mu'}\rho_{Y_2,2}^{\nu\nu'} \left[M^*_{\mu'\nu'}M_{\mu\nu}\right]_{Y_1,Y_2}}{Q_1^2 Q_2^2}\delta^{(4)}(q_1+q_2 - k)\;,
 \ee
 where 
 \be
 x_{1,2} = \frac{1}{\sqrt{s}}\left(E_{X} \pm p_{X,z}\right) = \frac{m_{X_\perp}}{\sqrt{s}} e^{\pm y_{X}}\;,
 \ee
 with $m_{X_\perp} = \sqrt{m_{X}^2 + p_{X_\perp}^2}$, while $-q_{i_\perp}$ is the transverse momentum imparted to proton beam $i$, such that $p_{X_\perp}=q_{1_\perp}+q_{2_\perp}$. Here $X$ ($=l^+ l^-$...) is the produced object, $\tilde{\beta}$ is as defined in~\cite{Harland-Lang:2019zur}, and the integral over $M_i^2$ in the photon density matrices are again understood as being performed simultaneously with the other phase space integrals.

Finally, we have discussed $Z$--boson initiated contributions above, but in general one can have contributions from initial--state $W$ bosons, via both the $W^+ W^- \to l^+ l^-$ and mixed $W^\pm  + q$ subprocesses. However, these will be further suppressed with respect to the $Z$--initiated channels. In particular, in the $W^+ W^- \to l^+ l^-$ case both initial-state particles must be $W$ bosons, and there can of course be no interference with $\gamma$--initiated diagrams. In the mixed case, again there is no $\gamma/W$ interference. This will generally lead to a strong suppression in these channels, and hence we do not consider them here for simplicity. However, these could be readily included within the SF framework.

\subsection{$\gamma/Z + q$ initiated production}\label{sec:zq}

\begin{figure}
\begin{center}
\subfigure[]{\includegraphics[scale=0.74]{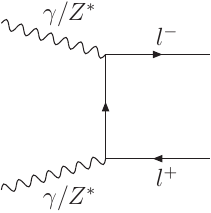}}\qquad
\subfigure[]{\includegraphics[scale=0.74]{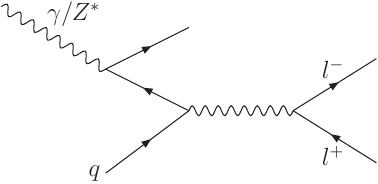}}\qquad
\subfigure[]{\includegraphics[scale=0.74]{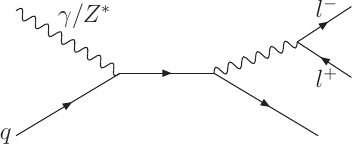}}
\caption{Representative Feynman diagrams for $\gamma/Z$ initiated lepton pair production.}
\label{fig:FDgam}
\end{center}
\end{figure}

In addition to pure $\gamma/Z$ initial states discussed above and shown in Fig.~\ref{fig:FDgam} (a), which features a $t$--channel lepton exchange, for the case of lepton pair production there are also mixed $\gamma/Z + q$ initiated contributions, as shown in Fig.~\ref{fig:FDgam} (b) and (c). These can readily be included within the same approach as above, i.e. with the corresponding initial state $\gamma/Z$ included via the SF formalism. However, as we will see the inclusion of these type of diagrams is necessarily more subtle, overlapping as it does with QED correction diagrams to the $q\overline{q}$ initiated process, and hence QED corrections to DGLAP evolution. Therefore, as well will see the benefit of including this type of diagram via the SF approach is less transparent.

Nonetheless, it is useful to explore how we might do so. To include these diagrams, we simply replace
\be
\rho_{Y_1,1}^{\mu\mu'}\rho_{Y_2,2}^{\nu\nu'} \left[M^*_{\mu'\nu'}M_{\mu\nu}\right]_{Y_1,Y_2} \to \frac{Q_2^2}{4\pi \alpha(Q_2^2)} \int\frac{{\rm d}M_2^2}{Q_2^2}\, \rho_{Y_1,1}^{\mu\mu'} \sigma_{\mu \mu'}^{2, Y_2}+\frac{Q_1^2}{4\pi \alpha(Q_1^2)}\int\frac{{\rm d}M_1^2}{Q_1^2} \,\rho_{Y_2,2}^{\mu\mu'} \sigma_{\mu \mu'}^{1, Y_1}\;.
\ee
Here, at LO we have
\be\label{eq:siggq}
\sigma_{\mu \mu'}^{i, Y_i} = \sum_{q,\overline{q}}  q(x_{B,i},\mu_F^2)  \left \langle A_{\mu} A_{\mu'}^* \right\rangle_{Y_i} \;,
\ee
where $A_\mu$ is the corresponding $\gamma/Z + q \to l^+ l^- + q$ amplitude, with a collinear initial--state quark/anti--quark carrying proton momentum fraction $x_{B,i}$. We include both photon and $Z$ initiated production of the final--state lepton pair in the amplitude, while the angle brackets indicate helicity/colour averaging as usual. The above result can be simply read off by matching with the corresponding LO QCD expression for the purely structure function type contribution. We use the $G_\mu$ EW scheme, with the PDG values~\cite{Zyla:2020zbs} for all corresponding inputs. 

Noting that away from the $Z$ peak region \eqref{eq:siggq} scales as $\sim 1/m_{ll}^2$, one can clearly see that these mixed contributions are suppressed by $\sim Q^2_i /m_{ll}^2$, and hence in the inclusive cross section should only play a minor role. On the other hand, near the $Z$ peak or in kinematic regions, such as at higher $p_\perp^{ll}$, where larger values of $Q^2_i$ are preferred and/or in the lower $m_{ll}$ region, the role of these contributions will be larger.

We also note the above expressions do not include any interference between the $\gamma\gamma$ and $\gamma + q$ diagrams. These can be effectively included at LO by simply reweighting the cross section according to
\be
R_{\rm intef.} = \frac{\sigma^{\rm LO}(\gamma_1 \gamma_2 + \gamma_1 q_2 + q_1\gamma_2)}{\sigma^{\rm LO}(\gamma_1\gamma_2) + \sigma^{\rm LO}(\gamma_1 q_2) +  \sigma^{\rm LO}(q_1 \gamma_2 )}\;,
\ee
where the $1,2$ subscript refers to the initiating beam, and the `LO' indicates that we calculate the corresponding LO quark--level diagrams, such that all initial--state photons are generated via the $q \to q \gamma$ process, i.e. by substituting the corresponding LO QCD expression for the structure functions in \eqref{eq:rho}. We have however checked that the impact of this interference is in general very small, and therefore given that the precise contribution will depend on the details of how we implement such effects (the above prescription is not unique), we do not include them in our results.

To make contact with the usual collinear approximation, we drop $Z$--initiated contributions in \eqref{eq:sighhvar} to give
 \be\label{eq:sighhf}
 \sigma = \frac{1}{2s}  \int  {\rm d}x_1 {\rm d}x_2\,{\rm d}^2 q_{1_\perp}{\rm d}^2 q_{2_\perp
} {\rm d \Gamma} \,\alpha(Q_1^2)\alpha(Q_2^2) \frac{1}{\tilde{\beta}}\frac{\rho_1^{\mu\mu'}\rho_2^{\nu\nu'} M^*_{\mu'\nu'}M_{\mu\nu}}{Q_1^2Q_2^2}\delta^{(4)}(q_1+q_2 - p_X)\;,
 \ee
We write $\tilde{\beta}$, defined in~\cite{Harland-Lang:2019zur}, as
 \be
 \tilde{\beta} = (1-\xi_1)(1-\xi_2) - \tilde{x}_1  \tilde{x}_2 \;,
 \ee
 where we have decomposed the photon momenta via
 \be
 q_1= \xi_1 p_1 + \tilde{x}_1 p_2 + q_{1_\perp}\;,\qquad q_2 = \tilde{x}_2 p_1 + \xi_2 p_2 + q_{2_\perp}\;.
 \ee
In the small $Q_i^2$ limit we have
\be
\left \langle A_{\mu} A_{\mu'}^* \right\rangle \approx R_{\mu \mu'}\sum_{\lambda_i} \left\langle  |A_{\lambda_i}|^2  \right\rangle \;,
\ee
where the sum is over the quasi--on--shell photon helicity $\lambda_i$, and $R$ is the metric tensor that is transverse to the photon and quark momenta, e.g. 
\be
R^{\mu\nu} \approx -g^{\mu\nu} + \frac{(p_1 q_2)(p_1^\mu q_2^\nu+p_1^\nu q_2^\mu)+Q_2^2 p_1^\mu p_1^\nu}{(p_1 q_2)^2}\;.
\ee
when the photon--initiator is from beam 2, and similarly in the alternative case. Here we have taken the  small $Q_i^2$ and high energy limit. Then we have
\be\label{eq:rhophot}
 \frac{1}{\alpha(\mu^2)}\int  \frac{{\rm d} Q^2}{Q^2}\alpha(Q^2)^2\rho_{\mu\nu}^i R^{\mu\nu} \approx  \frac{4\pi}{\xi_i}   f_{\gamma/p}^{\rm PF}(\xi_i,\mu^2)\;,
\ee
as in~\cite{Harland-Lang:2019eai}, where $\xi_i \geq x_i$ is the photon momentum fraction, and `PF' indicates this is the physical photon PDF, following the notation of~\cite{Manohar:2017eqh}, i.e.
\begin{align}\nonumber
  x f_{\gamma/p}^{\rm PF}(x,\mu^2) &= 
  \frac{1}{2\pi \alpha(\mu^2)} \!
  \int_x^1
  \frac{dz}{z}
  \int^{\frac{\mu^2}{1-z}}_{\frac{x^2 m_p^2}{1-z}} 
  \frac{dQ^2}{Q^2} \alpha^2(Q^2)
  \\  \label{eq:xfgamma-phys}
  &\cdot\Bigg[\!
  \left(
    zp_{\gamma q}(z)
    + \frac{2 x ^2 m_p^2}{Q^2}
  \right)\! F_2(x/z,Q^2)
    -z ^2
  F_L\!\left(\frac{x}{z},Q^2\right)
  \Bigg]\,.
\end{align}
A change of variables gives
\be
\frac{{\rm d}M_1^2}{Q_1^2}   {\rm d}^2 q_{1_\perp}= \frac{16\pi^3}{x_{B,1}}  (1-\xi_1)  {\rm d}\Gamma_{q'}\;.
\ee 
where $\Gamma_{q'}= {\rm d}^3 k_{q'} / [2 E_{q'} (2\pi)^3]$ is the phase space integral with respect to the outgoing quark and
\be
{\rm d}q_{2_\perp}^2 {\rm d}x_1 {\rm d}x_2 = \frac{\tilde{\beta}}{1-\xi_1} {\rm d}Q_2^2 {\rm d}\xi_1 {\rm d}\xi_2\;.
\ee
Combining the above we can then indeed write\footnote{For completeness we also clarify the original discussion on this point in the case of the $\gamma\gamma$--initiated cross section, as in Section 4 of~\cite{Harland-Lang:2019eai}. In particular, the relevant change of variables in this case is ${\rm d}q_{1_\perp}^2 {\rm d}q_{2_\perp}^2 {\rm d}x_1 {\rm d}x_2 = \tilde{\beta}  {\rm d}Q_1^2 {\rm d}Q_2^2 {\rm d}\xi_1 {\rm d}\xi_2$, which cancels the kinematic $\tilde{\beta}$ factor in \eqref{eq:sighhf}.}
\be\label{eq:gqpdf}
\sigma_{\gamma q} \approx\sum_{q,\overline{q}}  \int  {\rm d}\xi_1 {\rm d}\xi_2\,\left[q(\xi_1,\mu^2)f_{\gamma/p}^{\rm PF}(\xi_2,\mu^2)+f_{\gamma/p}^{\rm PF}(\xi_1,\mu^2)q(\xi_2,\mu^2)\right]\hat{\sigma}(\gamma +q \to l^+ l^- + q)\;,
\ee
where it is understood that the scale of the coupling in the $\gamma q$ subprocess should be taken as $\mu$ in order to match \eqref{eq:rhophot}.

Finally, in the PI case we might worry that for the class of diagrams in Fig.~\ref{fig:FDgam} (b)  there is an IR singularity associated with the region of phase space where the $\gamma \to q \overline{q}$ splitting becomes collinear. In particular, considering for simplicity the case that the initial--state photon comes from beam 2, in this limit \eqref{eq:gqpdf} becomes as usual:
\be\label{eq:gqpdfcoll}
\sigma_{\gamma q} \approx \sum_{q,\overline{q}}  \int  {\rm d}\xi_1 {\rm d}\xi_2\, q(\xi_1,\mu^2)\,\frac{\alpha}{2\pi}\int_0^{|k^2|_{\rm max}} \frac{{\rm d}|k^2|}{|k^2|} P_{q \gamma} \otimes f_{\gamma/p}^{\rm PF}(\xi_2,\mu^2)\hat{\sigma}(q\overline{q} \to l^+ l^-)\;,
\ee
where $P_{q\gamma}$ is the usual LO splitting function, $k$ is the momentum of the intermediate quark/anti--quark propagator, the sum over $q,\overline{q}$ is understood to match the $q\overline{q}$ initial state, and we use the shorthand
\be
g \otimes f(x) = \int_x^1 \frac{{\rm d}z}{z} f(z) g\left(\frac{x}{z}\right)\;.
\ee
We can see that the limit on the $k$ integral in \eqref{eq:gqpdfcoll} extends down to $k^2=0$ and hence we pick up the usual IR singularity that must be regulated and absorbed into the definition of the quark PDF in the usual way; the value of the upper kinematic limit is not important for our purposes, though it will be $\sim m_{ll}^2$. However, if we now consider the corresponding expression in the case of the full structure function calculation \eqref{eq:sighhf} then without writing down the rather more complicated general expression, we can simply observe that the relevant integral becomes:
\be\label{eq:kSF}
\int_0^{|k^2|_{\rm max}}  \frac{{\rm d}|k^2|}{|k^2|} \to \int_{Q^2_2}^{|k^2|_{\rm max}}  \frac{{\rm d}|k|^2}{|k|^2} = \ln\left(\frac{|k^2|_{\rm max}}{Q^2_2}\right)\;,
\ee
which is well defined, and certainly finite. In other words, the IR singularity that is present in the pure collinear calculation is automatically regulated in the SF calculation by the photon virtuality $Q_2^2$. Thus naively there is no requirement to absorb a $\gamma \to q/\overline{q}$ contribution into the quark/antiquark PDF, as there is in the collinear calculation. However, this is not in fact the case, as in the corresponding $q\overline{q}$ initiated processes, when we consider QED corrections we will pick up IR divergences from  $q \to q \gamma$ splittings that must be absorbed. These will generate as usual QED DGLAP evolution of the quark PDFs, and basic consistency requires that $\sim P_{q \gamma}$ contributions are also present. As well as being required by this, it is clear that \eqref{eq:kSF}, though finite, is logarithmically enhanced, and therefore may be more accurately resummed via DGLAP evolution, at least for larger scales $|k^2|_{\rm max}$.

This situation is rather close to the treatment of heavy quark flavours in the initial state, which while technically not requiring a treatment that goes beyond the fixed flavour scheme, often prefer the introduction of heavy quark PDFs in order to provide an accurate description at larger scales $Q^2 \gg m_q^2$. In the current case, given the smallness of the QED coupling $\alpha$, the phenomenological importance of this will be smaller, but consistency requires we follow a similar procedure, as discussed above. In analogy to the heavy quark case, we can proceed by including in our calculation the SF result, but  subtracting the logarithmic contribution as per \eqref{eq:kSF} in order to avoid double counting with the contribution that is generated from the QED DGLAP evolution of the quark PDFs. In more detail, we note that QED DGLAP evolution of the quark PDFs leads to
\be
q(x,\mu^2) = q(x,Q_0^2) + \frac{\alpha}{2\pi} \ln \left(\frac{\mu^2}{Q_0^2}\right)  P_{q \gamma}\otimes \gamma(x,\mu^2)+\cdots\;,
\ee
and similarly for the antiquark, to $O(\alpha)$. Here $Q_0$ corresponds to the input scale of the PDF set, which we take to be 1 GeV, as in~\cite{Harland-Lang:2019pla}. At $O(\alpha)$ we must then subtract the contribution
\be\label{eq:sub}
\sigma_{\gamma q}^{\rm sub} = \frac{\alpha}{2\pi} \ln\left(\frac{\mu^2}{Q_0^2}\right) \sum_{q,\overline{q}}  \int  {\rm d}\xi_1 {\rm d}\xi_2\, \left[q(\xi_1,\mu^2)\,P_{q \gamma}\otimes \gamma(\xi_2,\mu^2)+P_{q \gamma}\otimes \gamma(\xi_1,\mu^2)\,q(\xi_2,\mu^2)\,\right]\;,
\ee
from our structure function calculation, in order to avoid double counting with the contribution from the DGLAP evolution of the quark PDFs. This $\mu$ dependence in this result in particular matches the corresponding $\mu$ dependence of the quark PDFs, such that the combined $q\overline{q}$ and $q\gamma$ results maintain scale independence at LO in $\alpha$. Indeed, if one were to estimate the scale variation uncertainty on \eqref{eq:sub}, this would need to be done coherently with the $q\overline{q}$ channel. Scale variation in the $\gamma q$ contribution alone would overestimate the corresponding uncertainty. Nonetheless, it is the case that this introduces a scale variation uncertainty associated with the $\gamma q$ contribution, which is absent in the pure $\gamma\gamma$ case. 
Finally, there is in principle some uncertainty associated with the precise choice of $Q_0$, though this is also very small, and the value we take matches naturally that taken in~\cite{Harland-Lang:2019pla}.

We note that the generalisation of the above procedure to other processes is straightforward. In particular, we simply include a subtraction by folding the appropriate quark/antiquark initiated cross section with
\be
q_{\rm sub}(x,\mu^2) =\frac{\alpha}{2\pi} \ln \left(\frac{\mu^2}{Q_0^2}\right)  P_{q \gamma}\otimes \gamma(x,\mu^2)\;,
\ee
to $O(\alpha)$. If one were to include higher order QCD or EW corrections to the subprocess in the SF calculation, then the order of the above expression should simply be matched to this, provided the order of the DGLAP evolution applied in the non--SF contribution matches this. 

This subtraction should only be included for suitably inclusive cross section predictions, i.e. in the present case where the $q\overline{q}$ initial--state can contribute in its LO configuration. In particular, if we consider the dilepton transverse momentum distribution, then no double counting with the purely $q\overline{q}$ initiated process occurs at the order we consider, and the SF calculation provides the full prediction, with no subtraction required. In the end though, even in the case where technically one can argue as above that it should be included, we will see that the impact of this subtraction on the cross section is relatively minor in most (though not all) phenomenologically relevant regions.

In summary, one can certainly include such mixed diagrams via the SF procedure, but some care is needed to do this consistently.  In particular, one  needs to take care to suitably subtract any double counting with contributions generated by QED corrections to DGLAP in the pure QCD initial--state contributions. Even at lowest order this is not completely trivial, and certainly if one started to go to higher order in QCD, e.g. including gluon emission from the initial state quark lines in ~\ref{fig:FDgam} (b), then while the result would still be IR finite, the calculation would become more involved and the subtraction terms would be become more complicated again. Moreover, the formal precision of the SF calculation is clearly then tied to the order at which one does this subtraction. Thus, the benefits of using the SF approach for such a process are less clear, though there is no in principle obstacle to doing this. 

\section{Lepton pair production: results}

In the following sections we present some selected results for lepton pair production, as produced with the \texttt{SFGen} Monte Carlo generator. We will investigate in particular the impact of quark--initiated diagrams and initial--state $Z$ contributions by plotting both the pure $\gamma\gamma$ and the `total' cross section; the latter includes $Z$ and mixed $ \gamma/ Z + q$  contributions (including the corresponding subtraction described in Section~\ref{sec:zq} where relevant), in addition to the $\gamma\gamma$. We include the uncertainty due to the experimental uncertainty on the inputs, as discussed in Section~\ref{sec:SFgen}, for the total cross section, but omit this when showing the $\gamma\gamma$ result for demonstration purposes.

For comparison we also show the collinear $\gamma\gamma$ prediction at LO and in places NLO, using the \texttt{MMHT2015qed\_nnlo} photon PDF~\cite{Harland-Lang:2019pla}, with error bands corresponding to the PDF and factorization/renormalization scale variation uncertainty added in quadrature. For the collinear calculation we take $\mu_F=\mu_R$ and vary by the usual factor of 2 around the particular choice of $\mu_0$.
 In the SF calculation, there is a factorization scale uncertainty in the $\gamma/Z + q$ contribution, from the quark/antiquark PDF, see \eqref{eq:siggq}. For completeness we add this in quadrature to the experimental uncertainty discussed above in the total cross section, but as we will see this is in general a negligible source of uncertainty on the total cross section.

We will in places compare to the LO collinear predictions alone, as used in e.g.~\cite{Bertone:2017bme,Xie:2021equ}, in order to draw a qualitative comparison. However, the highest publicly available order we can calculate at corresponds to NLO, and this can naturally decrease the corresponding scale variation uncertainties and  improve the agreement with the SF result. Nonetheless, as explicitly demonstrated in ~\cite{Harland-Lang:2019eai} in the context of lepton--hadron scattering even at NLO the agreement with the SF approach is not perfect. Indeed, with this in mind we will in Section~\ref{sec:nlo} present a detailed comparison of the SF result with the highest available precision collinear calculation, namely at NLO. 

\subsection{Invariant mass distributions}\label{sec:invm}

\begin{figure}[h]
\begin{center}
\includegraphics[scale=0.64]{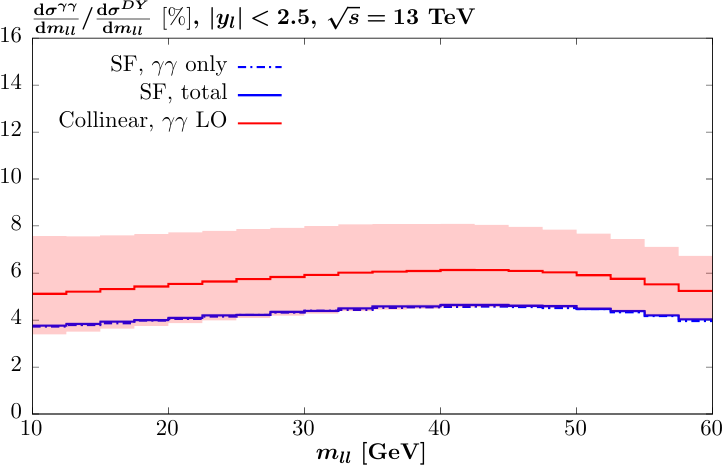}
\includegraphics[scale=0.64]{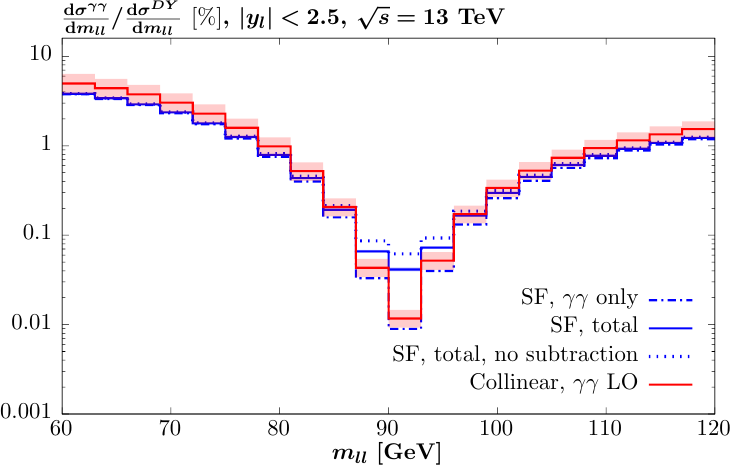}
\includegraphics[scale=0.64]{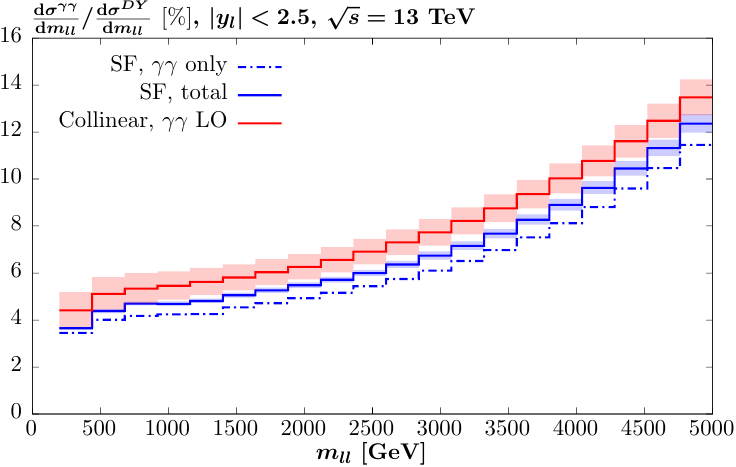}
\caption{Ratio of the photon--initiated cross sections for lepton pair production to the NLO QCD Drell--Yan cross section at the $13$ TeV LHC, as a function of the lepton pair invariant mass, $m_{ll}$.  The leptons are required to lie in the $|\eta_l|<2.5$ region. The LO collinear predictions and the structure function result are shown, in the latter case for both the pure $\gamma\gamma$ initiated and the total. In the former case the uncertainty band due to renormalization/factorization scale variation by a factor of two around the central value $\mu=m_{ll}$, added in quadrature with the much smaller PDF uncertainty, is given. In the latter case, the error band due to the experimental uncertainty on the structure functions and scale variation in the  $\gamma/Z + q$ component is shown added in quadrature (note in some regions this is comparable to the line width of the central value).}
\label{fig:mDY}
\end{center}
\end{figure}

We first consider the dilepton invariant mass distributions, shown in Fig.~\ref{fig:mDY}, taking the same mass regions and cuts as in~\cite{Harland-Lang:2019eai}. For the collinear predictions we take $\mu_0=m_{ll}$. Starting with the low mass region ($10 < m_{ll} < 60$ GeV) we can see that  the initial--state $Z$ contribution is negligible, as at these low masses the average $\gamma/Z$ virtuality $Q_{i}^2$ is rather low. In addition, for this inclusive observable we find that the $q \gamma$ contribution is equally very small.

Turning now to the intermediate mass region ($60 < m_{ll} < 120$ GeV), then away from the $Z$ peak the picture is rather similar to the low mass case, for the same reasons. Indeed, across the entire region the initial--state $Z$ contribution is again very small. However, in the $Z$ peak region we observe a significant enhancement with respect to the pure $\gamma\gamma$ that comes from the mixed $\gamma q$ contribution, where the dilepton pair can now be produced by a resonant intermediate $Z$ (see Figs.~\ref{fig:FDgam}). This provides the dominant contribution, as we might reasonably expect. We also show the result without applying the subtraction \eqref{eq:sub}, to assess its impact, and can see that this results in a non--negligible reduction in the cross section in the $Z$ peak region, while outside of this the effect is tiny. Broadly, it is clear that even the total cross section, while enhanced with respect to the $\gamma\gamma$, still constitutes a very small fraction of the DY cross section, representing as it does as genuine NLO EW correction.

We next consider the high mass ($120 < m_{ll} < 5000$ GeV) region\footnote{We note that minor errors in the implementation of both the collinear and SF calculations of lepton pair production have been found in the results of~\cite{Harland-Lang:2019eai}, subsequent to publication. The impact of these is generally very small, but becomes more visible at  high masses: hence one can observe that both of these contributions differ with respect to the previously published results at higher mass. This issue in addition affected the evaluation of the experimental uncertainty on the SF result, which is now seen to be somewhat larger at high mass.}. We can see that as in~\cite{Harland-Lang:2019eai} the pure $\gamma\gamma$ SF contribution lies somewhat below the collinear prediction, outside the scale variation uncertainty bands of the latter. The total SF cross section is enhanced with respect to the $\gamma\gamma$, in particular at higher masses, due to the impact of initial--state $Z$ contributions, which play more of a role as the average $\gamma/Z$ virtuality $Q^2$ increases in this region. Clearly though this is not a question of improving the agreement with the collinear $\gamma\gamma$ calculation, which explicitly does not include such a contribution. 

Finally, we can see that the experimental uncertainty on the SF calculation is very small in the low and intermediate mass regions (though shown, it is rather less than the line thickness), while at the highest masses it is a little larger though still small. In more detail, this varies from $\sim 1-3\%$ across the mass region, with the scale variation uncertainty in the $\gamma/Z + q$ contribution being a negligible component of this. 

In all cases the LO collinear prediction tends to lie above the SF result, though at lower and intermediate masses the result is in  agreement within the large scale variation uncertainties. However, as we have seen at higher mass even then the result lies above the SF prediction. We now investigate the extent to which this situation improves upon the inclusion of NLO corrections to the collinear calculation.

\subsection{Collinear predictions: the impact of NLO corrections}\label{sec:nlo}

\begin{figure}[h]
\begin{center}
\includegraphics[scale=0.64]{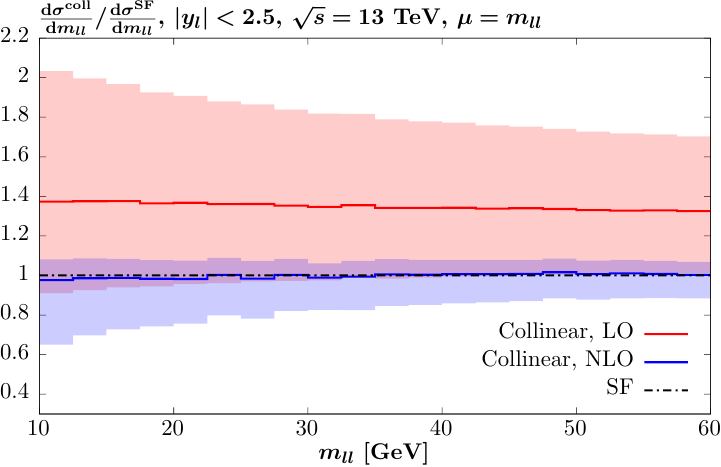}
\includegraphics[scale=0.64]{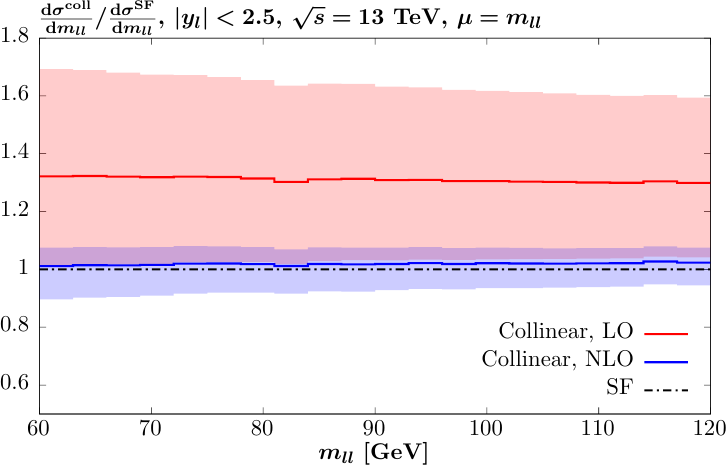}
\includegraphics[scale=0.64]{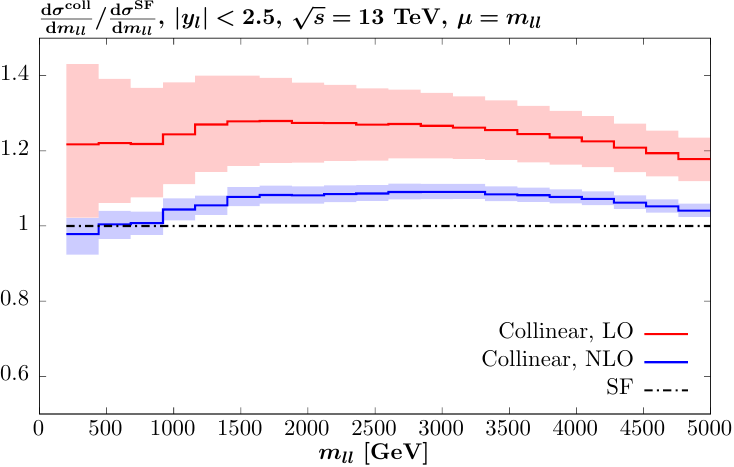}
\caption{Ratio of the LO and NLO collinear predictions to the structure function result, for photon--initiated lepton pair production, as a function of the lepton pair invariant mass, $m_{ll}$. The leptons are required to lie in the $|\eta_l|<2.5$ region. In the collinear case the uncertainty band due to renormalization/factorization scale variation by a factor of two around the central value $\mu=m_{ll}$ is shown.}
\label{fig:mDY_nlo_rat}
\end{center}
\end{figure}

In this section we consider the impact of NLO corrections on the collinear calculation for PI production, and compare with the SF calculation. We note that while the fixed--order calculation of the purely QCD contribution to lepton pair production has been performed with very high precision, up to even ${\rm N}^{3}$LO~\cite{Duhr:2020seh,Duhr:2020sdp,Chen:2021vtu}, for the PI component NLO represents the highest publicly available result. Hence, we will be comparing with the highest precision calculation of PI production in collinear factorization that is currently available.

At NLO in the collinear calculation, as well as the diagrams shown in Figs.~\ref{fig:FDgam} (b) and (c), the diagram of the topology (a), but where one of the initial--state photons is produced via a $q \to q \gamma$ splitting, enters. As discussed in Section~\ref{sec:zq}, away from the $Z$ peak the (b) and (c) topology diagrams are kinematically suppressed with respect to the $t$--channel diagram (a), i.e. due to $t$--channel lepton exchange. Therefore, while these are certainly amenable to being included in a collinear calculation, and indeed can be, for simplicity we choose to focus in this section purely on the topology of diagram (a), along with its corresponding $q \to q \gamma$  NLO contribution. This is of course not strictly the entire NLO contribution, even if away from the $Z$ peak it is by far the dominant one, but for our purposes, i.e. to evaluate the extent to which including  NLO corrections stabilises the scale variation of the collinear calculation of diagram (a), we will for clarity only consider this contribution, omitting diagrams (b) and (c) entirely. 

In order to achieve this, we make use \texttt{MadGraph5\_aMC@NLO}~\cite{Alwall:2014hca,Frederix:2018nkq}, suitably modified in order to remove the diagrams of type (b) and (c), as well as those due to final--state photon emission and virtual EW corrections to the $\gamma\gamma \to l^+l^-$ process, which are not relevant for the current comparison. We in addition correct the results in order to include a running value of $\alpha$, which is not applied at present in the public version of \texttt{MadGraph5\_aMC@NLO}; the collinear photon PDF is extracted under the assumption that one will use a running $\alpha$ and hence this is required to ensure overall consistency, and indeed genuine NLO accuracy. 

In Fig.~\ref{fig:mDY_nlo_rat} we show the same invariant mass distribution as in the previous section, but now plotting the ratio of the collinear prediction at LO and NLO to the SF result. In the collinear case we show the uncertainty band due to varying the default renormalization/factorization scale $\mu_0= m_{ll}$ by a factor of 2, while in the latter case we only include the pure $\gamma\gamma$ contribution, i.e. again the $t$--channel topology of Fig.~\ref{fig:FDgam} (a), so that we are comparing completely like--for--like. In the LO case this therefore simply comes from taking the ratio of the  corresponding curves in Fig.~\ref{fig:mDY}, although we do not include any uncertainty on the SF calculation. This can be read off instead from Fig.~\ref{fig:mDY}, though it should be emphasised that this is largely correlated with the corresponding PDF uncertainty in the collinear case, and hence would mostly cancel in the ratios we show here. Moreover, these absolute size of these uncertainties is we recall at the percent level across most of the mass region.

Starting with the low mass region ($10 < m_{ll} < 60$ GeV), we can see that broadly speaking the inclusion of NLO corrections indeed leads to smaller scale variation uncertainties and to an improved agreement with the SF result. However, the scale variation uncertainty is  not negligible: in the lowest mass bin it is over $40\%$ in total (i.e. from the lowest to highest value with respect to the central), decreasing to $\sim 20\%$ in the highest mass bin. This is considerably larger than the PDF uncertainty,  and indeed the experimental uncertainty on the SF calculation,  which both enter at the level of $\sim 1\%$ in this region. We note that the rather precise level of agreement of the central collinear prediction with the SF is certainly accidental, given the size of the scale variation band; indeed, we will show below that once a cut is imposed on the lepton $p_\perp$ this precise level of agreement is not maintained. This trend essentially continues for the intermediate mass region ($60 < m_{ll} < 120$ GeV). That is, we find an improved agreement at NLO, with a reduced scale variation band, but which is nonetheless significantly larger than the PDF uncertainty. The total scale variation band in particular ranges from $\sim 20-10\%$ with increasing mass, while the PDF uncertainty is again at the $\sim 1\%$ level. However, it should be noted that as we only include the $t$--channel topology of Fig.~\ref{fig:FDgam} (a), this contribution is of limited phenomenological relevance as we approach the $Z$ peak region.

\begin{figure}
\begin{center}
\includegraphics[scale=0.64]{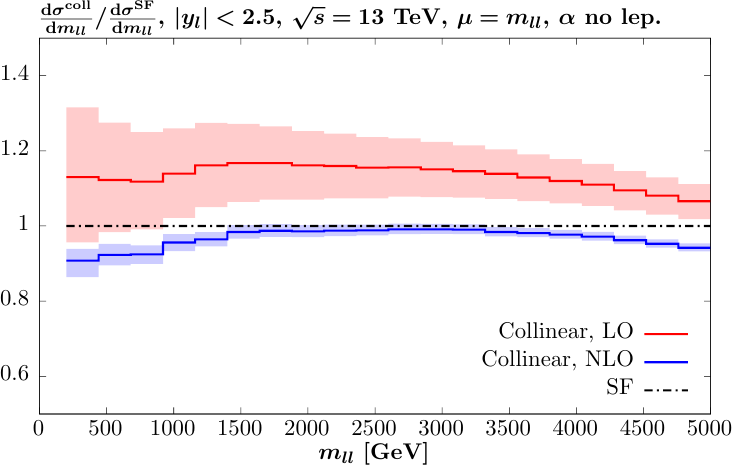}
\caption{As in Fig.~\ref{fig:mDY_nlo_rat}, but for high mass region alone, and with no leptonic loops included in the running of $\alpha$.}
\label{fig:mDY_nlo_nolep}
\end{center}
\end{figure}

Turning now to the high mass ($120 < m_{ll} < 5000$ GeV) region, we observe the same basic trend, however interestingly we find that the NLO collinear result does not agree with the SF within its scale variation band. The same trend was observed in the previous section for the LO result, as is clear here as well. A potential cause of this may lie in the fact that the \texttt{MMHT2015qed\_nnlo} PDFs we use to evaluate the collinear prediction do not include leptonic contributions in the virtual photon splitting function $P_{\gamma\gamma}$. The argument for doing this rests on the fact that lepton PDFs are not included in this set, and hence including such lepton loops in $P_{\gamma\gamma}$ would lead to a (albeit rather small) violation in the momentum sum rule, see~\cite{Harland-Lang:2019pla} for discussion. However, it is clear from the discussion in~\cite{Manohar:2016nzj,Harland-Lang:2019eai} that this effectively corresponds to evaluating the overall factor of $1/\alpha(\mu^2)$ in the definition of the photon PDF, see \eqref{eq:rhophot}, without leptonic loop contributions to the running of $\alpha$. There is therefore in principle a mismatch occurring in our results, which as usual do include such leptonic contributions in the running of $\alpha$. The effect of this is rather small until the highest invariant masses, where the evolution length of $\alpha$ is increased, and the difference is larger, increasing the resultant PI cross section by up to $\sim 10\%$. To demonstrate this, in Fig.~\ref{fig:mDY_nlo_nolep} we show the same comparison as before, but now only including quark contributions to the running of $\alpha$\footnote{In principle there are of course $W$ loop contributions, but  these are not included in the photon PDF via $P_{\gamma\gamma}$, and we therefore omit them in the running of $\alpha$.}. We can see that the agreement is indeed improved, although the LO generally still lies outside the SF within scale variation bands, and there are still regions where at NLO the collinear result disagrees within the scale variation bands.  In the highest mass bins, this may be due to the increased sensitivity to the $z\to 1$ region of the integral entering the photon PDF, see \eqref{eq:rhophot}, and in particular in the upper limit of the $Q^2$ region, which could lead to reduced convergence in the collinear result. In the future, it would be interesting to investigate the comparison using other photon PDF sets~\cite{Manohar:2016nzj,Bertone:2017bme,Xie:2021equ} which do including the leptonic contribution to  $P_{\gamma\gamma}$, and hence should be more consistent with the default running of $\alpha$. Indeed, as observed in Fig.~(24) of~\cite{Harland-Lang:2019pla}, the impact of including lepton loops in $P_{\gamma\gamma}$ is dependent on $x$ as well as scale, and this will not be accounted for in the simple correction applied above. In terms of the size of the scale variation uncertainty, this is at the level of a few percent in this high mass region, and hence is of a similar size to, or even  at the highest masses smaller than,  the PDF uncertainty. However, as we have seen the difference with respect to the SF baseline is in places larger than that.

\begin{figure}
\begin{center}
\includegraphics[scale=0.64]{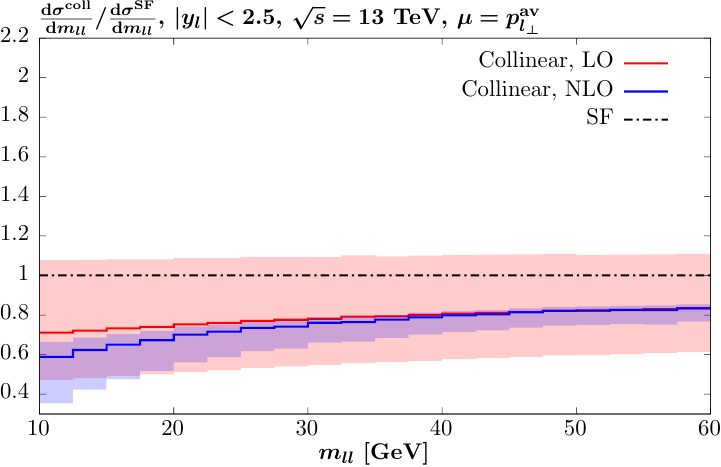}
\includegraphics[scale=0.64]{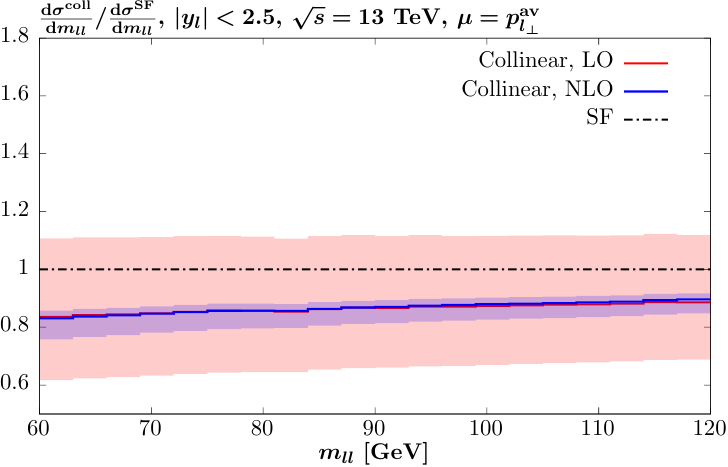}
\includegraphics[scale=0.64]{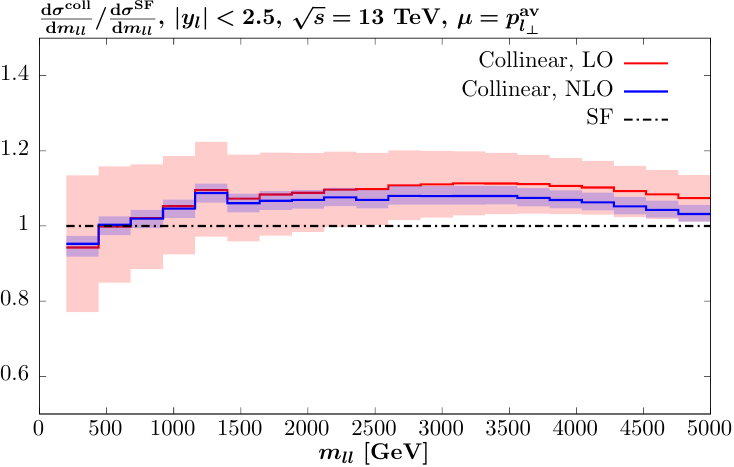}
\caption{As in Fig.~\ref{fig:mDY_nlo_rat}, but with the renormalization/factorization scale $\mu$ taken as the scalar average of the lepton transverse momentum $p_\perp^{\rm av}$.}
\label{fig:mDY_nlo_rat_pt}
\end{center}
\end{figure}

So far we have taken the dilepton invariant mass, $m_{ll}$, as the central scale choice. Another possible choice is the lepton transverse momentum, and we therefore investigate the impact of taking this instead. For concreteness we take the scalar average of the lepton transverse momenta, but have checked that for example taking the leading lepton $p_\perp$ gives a rather similar result, and in fact  increases the difference with respect to the SF calculation we observe at NLO. In Fig.~\ref{fig:mDY_nlo_rat_pt}
we show the same comparisons as before, but now with this choice of scale. For LO kinematics we have $p_{\perp}^l \leq m_{ll}/2$, and hence as expected at LO the result is rather lower than the $\mu = m_{ll}$ case. We might naively expect the NLO calculation to stabilise the result around the SF prediction, as was observed for this previous choice of scale, but this is not the case for the low and intermediate mass regions. While the scale variation bands are smaller, and the NLO correction itself rather small with respect to the central LO prediction, the result is systematically lower than the full SF calculation, and well outside these scale variation bands. The reason for this appears to lie in the correlation between the lepton transverse momenta and the transverse momentum of the off--shell initiating photon, due to the $q\to q \gamma$ splitting in the NLO diagram. This can lead to values of the $p_\perp^l$ that are rather higher than the LO kinematic limit  $p_{\perp}^l \leq m_{ll}/2$, and this appears to reduce the convergence of the collinear result at lower masses. In the high mass region, the impact of this effect is relatively smaller, and we observe greater stability in the result, and consistency with the $\mu = m_{ll}$ case. Thus this scale choice appears to be rather disfavoured. However, it should be emphasised that we can only reach this conclusion after directly performing the calculation in the SF approach, where there is no such ambiguity due to the choice of scale. Without such a comparison, all we could conclude is that the two scale choices lead to rather differing results until fairly high mass, which at NLO largely do not overlap in their scale uncertainty bands. 

\begin{figure}
\begin{center}
\includegraphics[scale=0.64]{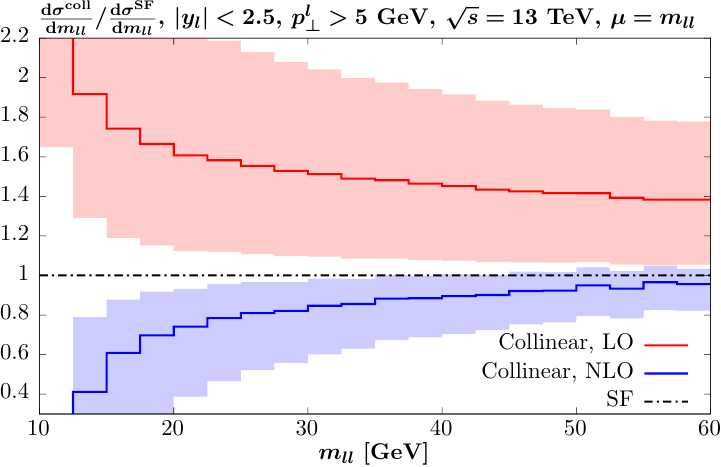}
\includegraphics[scale=0.64]{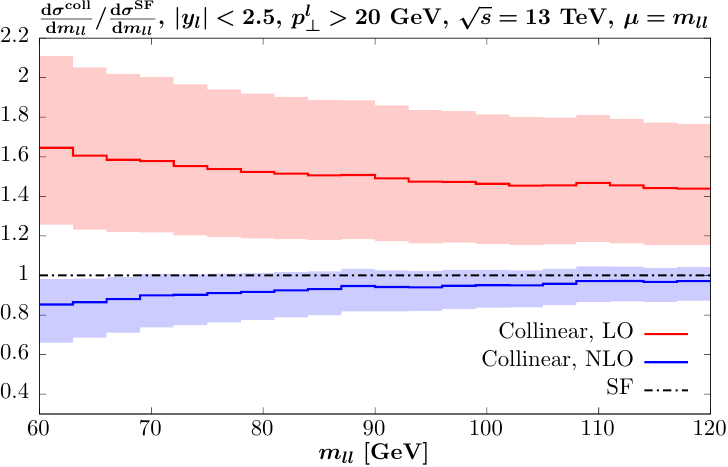}
\includegraphics[scale=0.64]{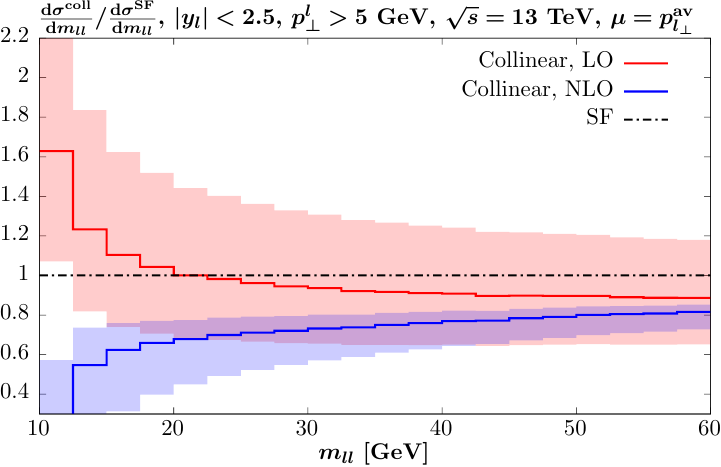}
\includegraphics[scale=0.64]{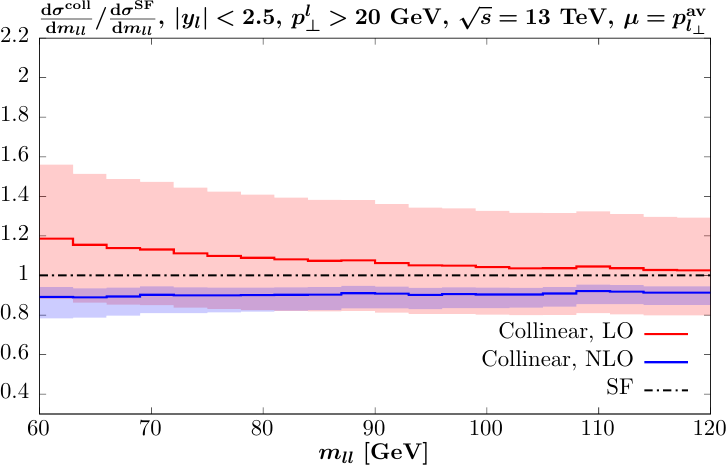}
\caption{As in Fig.~\ref{fig:mDY_nlo_rat}, but for the low and intermediate mass ranges alone, and with a cut on the lepton transverse momenta of $p_\perp^l > 5$ GeV and 20 GeV imposed, respectively. The leptons are as before required to lie in the $|\eta_l|<2.5$ region.The upper (lower) panel corresponds to $\mu=m_{ll}$ ($p_\perp^{\rm av}$).}
\label{fig:mDY_nlo_rat_ptc}
\end{center}
\end{figure}

Finally, we might worry that the above results are all produced with no cut imposed on the lepton $p_\perp$, which is certainly not experimentally relevant, and could influence the comparison. We therefore show in Fig.~\ref{fig:mDY_nlo_rat_ptc} the low and intermediate mass regions again, but now imposing a cut of  $p_\perp^l > 5$ (20) GeV on the lepton transverse momentum in the low (intermediate) cases; at high mass the impact of an equivalent $O(10)$ GeV cut would be less visible, and hence we do not show this here for brevity. The impact of this is relatively moderate, but not negligible. We find in general that the agreement between the NLO collinear and the SF for $\mu=m_{ll}$ has deteriorated, with the SF result lying on the edge of the scale uncertainty band or even beyond it. For  $\mu=p_\perp^{\rm av}$  the difference with respect to the no $p_\perp^l$ cut case is somewhat smaller. However, in both cases the scale variation band becomes rather larger as one approaches the lower mass region, where the impact of the cuts is more relevant. The agreement between the two scale choices is in addition somewhat improved, although not the agreement with the full SF result. 

In summary, we have found that when we take $\mu=m_{ll}$ as the central scale choice, the impact of NLO corrections is broadly to improve the matching of the collinear prediction to the full SF result for the $t$--channel topology of diagram (a) in Fig.~\ref{fig:FDgam}, with reduced scale variation uncertainties, in comparison to the LO calculation. Nonetheless at low to intermediate masses, $m_{ll}\lesssim 500$ GeV, the scale variation is still $O(10\%)$ or more, and hence is significantly larger than the PDF uncertainty. At higher masses, $m_{ll}\gtrsim 500$ GeV, the scale variation uncertainty is rather smaller, at a similar level to the PDF uncertainties, and the matching between the NLO and the SF relatively good, once we account for an apparent mismatch in the treatment of the running of $\alpha$ and the evolution of the \texttt{MMHT2015qed\_nnlo} PDFs. After correcting for this the agreement is improved, but it is not always within scale variation bands. On the other hand, once we impose realistic cuts on the lepton transverse momenta the agreement in the low to intermediate mass region deteriorates somewhat. Moreover, if we instead take a scale  $\mu=p_\perp^{\rm av}$, given in terms of the lepton $p_\perp^l$, the agreement between the NLO result and the SF is poor, and well outside the scale variation bands. Therefore, while the stability and the precision of the prediction can be improved by the inclusion of the NLO corrections, this is not guaranteed, and the remaining scale variation uncertainties can remain significantly larger than the percent level PDF uncertainties, as was the case at LO. We re--emphasise here that for this $t$--channel PI process, which kinematically dominates away from the $Z$ peak, the SF result has no such scale variation uncertainty, or indeed ambiguity due to the choice of dynamical scale, and therefore this issue is bypassed entirely. Indeed, it is only by comparing with the SF result that we can say which scale choice,  $\mu=m_{ll}$ or $\mu=p_\perp^{\rm av}$, is preferred. The NLO calculation on the other hand consists in explicitly re--calculating the $q\to q\gamma$ splitting contribution to the PI process, which is already implicitly included in the SF result, but in a manner that introduces a scale ambiguity, and hence source of uncertainty, that is not present in the SF case.

\subsection{Dilepton $p_\perp^{ll}$ distribution: revisited}\label{sec:dileppt}

We now consider the impact on the dilepton transverse momentum, $p_\perp^{ll}$, distribution. As in~\cite{Harland-Lang:2019eai}, we consider the same event selection as the ATLAS 8 TeV measurement~\cite{Aad:2015auj}, which is presented both on and off the $Z$ peak. The leptons are required to have $p_\perp^l >20$ GeV and $|\eta_l|<2.4$. We now compare to the on--peak region, as well as considering the impact of the $\gamma/Z + q$ and initial--state $Z$ contributions.

\begin{figure}
\begin{center}
\includegraphics[scale=0.655]{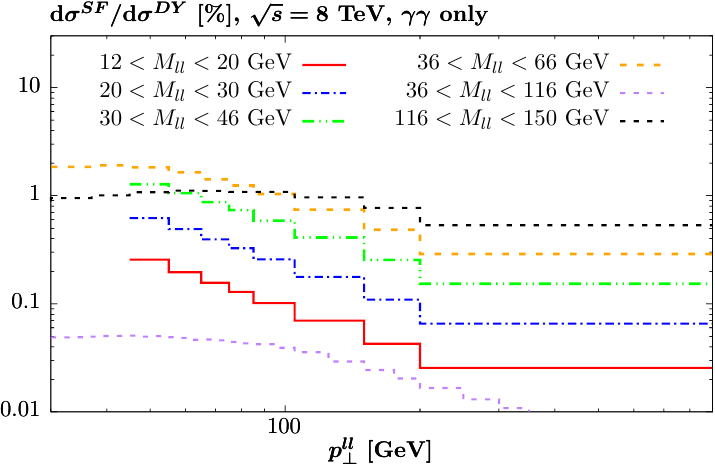}
\includegraphics[scale=0.655]{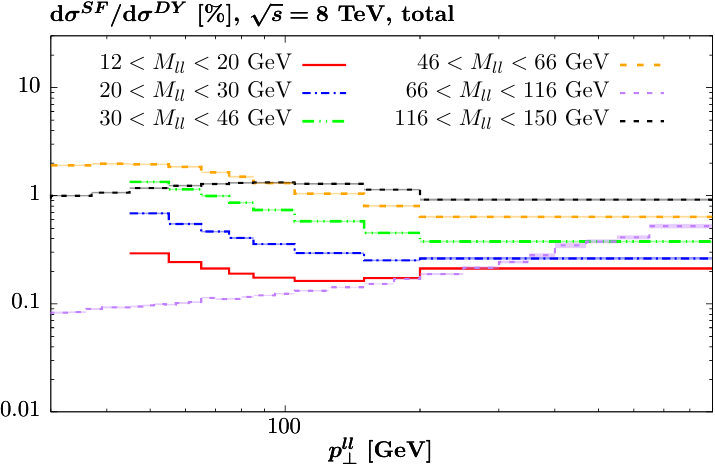}
\caption{Percentage contribution from photon--initiated production to the lepton pair $p_\perp$ distribution, within the ATLAS~\cite{Aad:2015auj} off--peak event selection, at 8 TeV. The pure $\gamma\gamma$  and total SF results are shown in the left and right plots, respectively. The error band due to the experimental uncertainty on the structure functions and scale variation in the  $\gamma/Z + q$ component is shown (note in some regions this is comparable to the line width of the central value). The QCD predictions correspond to NNLO QCD theory~\cite{Bizon:2018foh}.}
\label{fig:theorypt}
\end{center}
\end{figure}

In Fig.~\ref{fig:theorypt} we show results in the $p_\perp^{ll} > 30$ GeV region, and consider the ratio to NNLO QCD theory, produced with \texttt{NNLOjet}~\cite{Bizon:2018foh}. We do not include EW corrections to the QCD process, but note these are found to reduce the cross section at high $p_\perp^{ll}$ (see e.g.~\cite{Kallweit:2015fta}) and hence will increase the relative contribution from PI production somewhat in this region. In the left figure we show the pure $\gamma\gamma$ SF predictions, where these contributions are  seen to be small, but not necessarily negligible, being at the percent level at lower  $p_\perp^{ll}$ in some mass regions. In the on--peak region the relative contribution is as expected lowest. Now, in the right hand plot we show the total SF cross section, and we can see that this leads to a sizeable increase in the cross section at higher $p_\perp^{ll}$, while in the lower $p_\perp^{ll}$ region the results are relatively stable. This is due both to the $\gamma/Z +q$ and initial--state $Z$ contributions. The former component plays a significant role due to the $\sim Q_i^2/m_{ll}^2 \sim (p_\perp^{ll})^2/m_{ll}^2$ scaling discussed in Section~\ref{sec:zq}, and is particularly large at lower $m_{ll}$, for the same reason. The latter component leads to an enhanced at larger $p_\perp^{ll}$ across all mass regions, due to the larger $Q_i^2$ probed in this region; in the largest $p_\perp^{ll}$ bin it increases the cross section by a factor of $\sim 2$. For the $Z$ peak region, the major contribution is simply the inclusion of the $\gamma/Z+ q$ component, which then allows resonant production of the dilepton system, with the relative contribution being largest at larger $p_\perp^{ll}$ (though note that the absolute cross section is of course steeply falling).

\begin{figure}
\begin{center}
\includegraphics[scale=0.655]{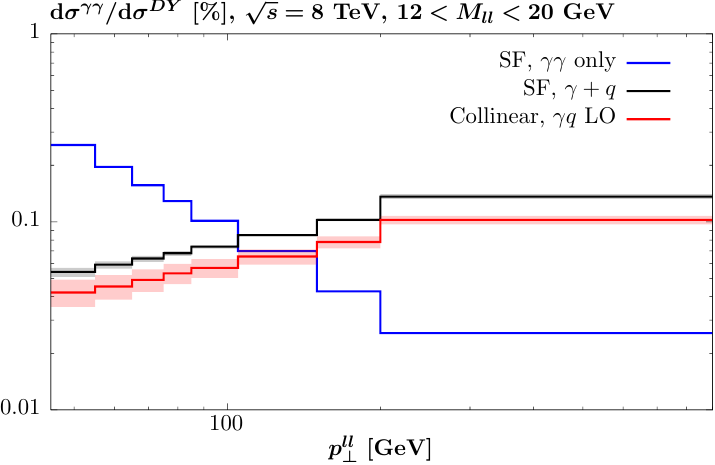}
\includegraphics[scale=0.655]{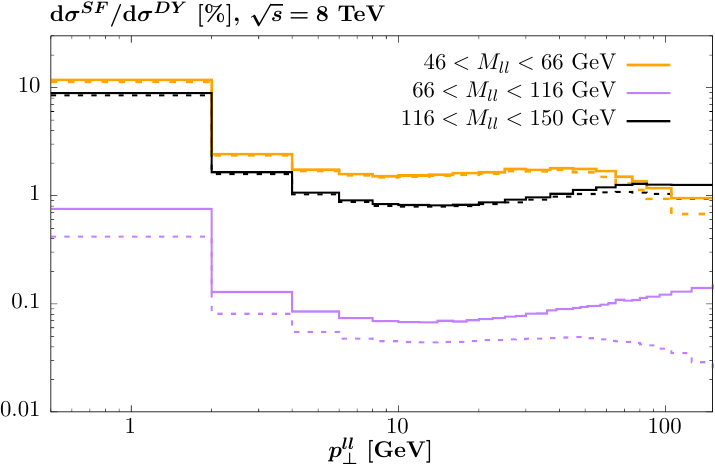}
\caption{Percentage contribution from photon--initiated production to the lepton pair $p_\perp$ distribution, within the ATLAS~\cite{Aad:2015auj} off--peak event selection, at 8 TeV. The QCD predictions correspond to NNLO + NNLL QCD theory~\cite{Bizon:2018foh}.  The left plot corresponds to the $12 < m_{ll} < 20$ GeV mass region, and shows the $\gamma\gamma$ and $\gamma + q$ SF contributions, as well as the collinear $\gamma + q$. In the collinear case the uncertainty band due to  scale variation by a factor of two around the central value $\mu_0=m_{ll}$, is given. The left plots shows the total (solid line) and pure $\gamma\gamma$ (dashed) contributions for different mass regions. For both plots, in the total SF case, the error band due to the experimental uncertainty on the structure functions and scale variation in the  $\gamma + q$ component is shown.}
\label{fig:theorylowpt}
\end{center}
\end{figure}

In more detail, in Fig.~\ref{fig:theorylowpt} we show the $12< m_{ll} < 20$ GeV mass region of Fig.~\ref{fig:theorypt}, but now focussing on the relative impact of the $\gamma + q$ and $\gamma\gamma$ contributions; we omit the initial--state $Z$ contribution here for clarity. We can see that for the SF results, the trend is as expected, namely at lower $p_{\perp}^{ll}$ the $\gamma\gamma$ contribution dominates, whereas at higher $p_{\perp}^{ll}$ the $\gamma + q$ takes over. The small uncertainty band in the latter case is dominated by the scale variation uncertainty on this mixed contribution due to the initial--state quark/antiquark, with the experimental uncertainty on the pure $\gamma\gamma$ case being as usual very small, and indeed not visible on the plot if it were included.
We in addition however show the $ \gamma + q$ contribution as calculated in the purely collinear LO case, that is via quark/antiquark and photon PDFs; due to the $p_{\perp}^{ll}> 45$ GeV cut that is imposed in this mass region this cross section is free of initial--state collinear singularities. The scale variation uncertainty about $\mu_0 = \sqrt{m_{ll}^2 + (p_{\perp}^{ll})^2}$ is shown, and is significantly larger, due to the introduction of a photon PDF in this case. Interestingly, this lies below the SF prediction, in particular at larger $p_{\perp}^{ll}$. This is as we might expect, given that a non--zero photon $q_\perp$ is permitted in the SF case, but not in the LO collinear result. 

Finally, in Fig.~\ref{fig:theorylowpt} (right) we focus on the low $p_{\perp}^{ll}$ region, considering the ratio to NNLO+N${}^3$LL resummed QCD predictions produced with \texttt{NNLOjet+RadISH}~\cite{Bizon:2018foh}. We consider three mass regions, both and on and off $Z$ peak, and show both the pure $\gamma\gamma$ (dashed line) and total (solid line) SF predictions. Away from the $Z$ peak, the former component is completely dominant until we approach larger $p_{\perp}^{ll}$ values. On the $Z$ peak, as usual the inclusion of the $\gamma + q$ component allows for resonant dilepton production, and hence a significant enhancement. Of particular interest though is the significant enhancement observed in the pure $\gamma\gamma$ contribution in the lower $0 < p_\perp^{ll} < 2$ GeV mass region. As discussed in~\cite{Harland-Lang:2019eai}, this is explained in part by the Sudakov suppression in the QCD contribution in this region, which is absent in the $\gamma\gamma$ channel. However, another key factor in this is that the $\gamma\gamma$ cross section is particularly peaked in this region, due to the significant contribution from elastic photon emission. It is important to point out here that this is an element of the cross section that by construction can never be accounted for within a purely collinear photon PDF framework. By including higher orders in the collinear calculation one can improve the accuracy from the $Q^2$ continuum region, and in the present case account for the $p_\perp^{ll}$ dependence of the cross section due to this, which is trivial at LO. However this is not the case for the elastic component, or indeed the low $Q^2$ resonant and non--resonant components, all of which cannot be modelled differentially in this way, and are the relevant components in the low $p_\perp^{ll}$ region. Thus the only way to precisely account for this low $Q^2$ (and hence low $p_\perp^{ll}$) region is via the SF result\footnote{One could in principle also do this via a $k_\perp$ dependent unintegrated photon PDF, though there is no motivation for doing this in favour of the SF result, see~\cite{Harland-Lang:2019eai} for further discussion.}. 

A potentially significant impact of the above discussion is that, while in the inclusive cross section the $\gamma\gamma$ contribution to the on--peak dilepton cross section is strongly suppressed, this is in part counterbalanced at very low $p_\perp^{ll}$ by the enhancement in the $\gamma\gamma$, and corresponding Sudakov suppression in the QCD DY cross sections. In the lowest mass bin, the $\gamma\gamma$ contribution is at the $\sim 4$ per mille level. Of course this is still rather small, but it is worth recalling that such a level of precision, or even better, is required in the LHC high precision programme. This could in particular be relevant for extractions of the $W$ boson mass, as we will now discuss.

\subsection{Impact on $M_W$ determination}\label{sec:mw}

As is well known, while the $W$ boson mass cannot be determined via its leptonic decay directly due to the invisible neutrino in the final state, one can construct kinematic observables that are sensitive to $M_W$. The two most common such variables are the transverse momentum of the charged decay lepton, $p_{l_\perp}$, and the $W$ boson transverse mass 
\be
m_\perp = \sqrt{2 p_{l_\perp} p_{l_\perp}^{\rm miss} (1- \cos \Delta \phi)}\;,
\ee
where $\Delta \phi$ is the azimuthal opening angle between the charged lepton and the missing transverse momentum. In particular, both of these variables exhibit Jacobian peak structures that are sensitive to the value of $M_W$, and therefore by fitting the measured distributions with a range of templates for differing mass values, one can extract information about $M_W$. Indeed, this was precisely the technique used by the ATLAS 7 TeV analysis~\cite{Aaboud:2017svj}, for which the corresponding total uncertainty of $19$ MeV is comparable to previous Tevatron determinations, and is starting to bear down on indirect constraints. An important aim for the future LHC and HL--LHC programme is to reduce this uncertainty further, see~\cite{Azzi:2019yne}.

To achieve this level of precision in a hadron collider environment, a precise and accurate control over the underlying $W$ production process is essential. The impact of PDF uncertainties is particularly important in this context~\cite{Bozzi:2011ww,Bozzi:2015hha,Bagnaschi:2019mzi,Hussein:2019kqx,Farry:2019rfg}, but there are numerous other sources, see~\cite{CarloniCalame:2016ouw} for a review. One possible source of contamination that has however not so far been considered in detail is the contribution from PI production. At first glance this might appear completely irrelevant, given $l \nu$ production can of course not occur directly from the $\gamma\gamma$ initial state. However, the issue here is that to achieve high precision it is necessary to perform a systematic tuning of certain theoretical inputs to $Z/\gamma\to l^+ l^-$ production, as  discussed in detail in Section 6 of~\cite{Aaboud:2017svj}. One particularly relevant case in point is the tuning of certain QCD parameters in \texttt{PYTHIA 8}~\cite{Sjostrand:2007gs}, relating e.g. to the intrinsic partonic $k_\perp$, to the measured dilepton $p_\perp^{ll}$ distribution; these are then used as an input in the $W$ mass measurement. Thus any contribution from $\gamma\gamma$--initiated production in the $Z/\gamma$ case, precisely because it is entirely absent in the $W$ case\footnote{We focus on the pure $\gamma\gamma$ initial state for this reason, as it should be the largest source of difference. The mixed $\gamma + q$ contributions will be present in both the $W$ and $Z/\gamma$ channels, but here the effect more naturally forms one part of the more general issue of EW corrections, which of have been studied in detail already and are not the focus here.},  may bias the results if it is not accounted for correctly, by being tuned away in the comparison with the dilepton data. In the ATLAS analysis~\cite{Aaboud:2017svj}, tuning is for example performed to a measurement of the $Z$ boson transverse momentum distribution presented in~\cite{Aad:2014xaa}. Here, the PI contributions is subtracted using the collinear LO predictions with the now outdated \texttt{MRST2004QED} PDF set~\cite{Martin:2004dh}.

\begin{figure}
\begin{center}
\includegraphics[scale=0.64]{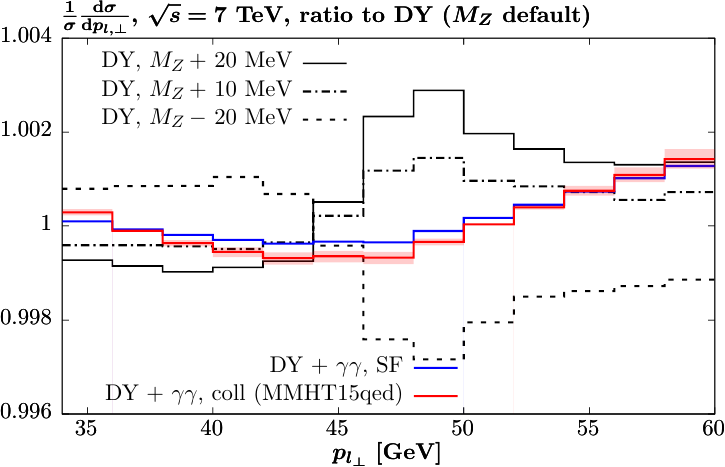}
\includegraphics[scale=0.64]{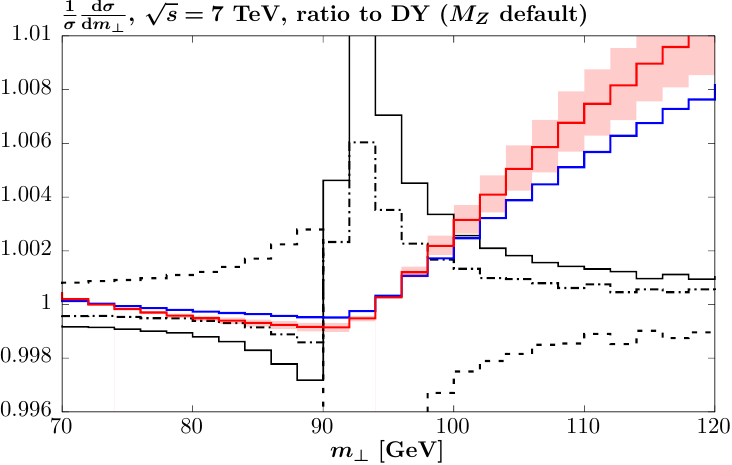}
\caption{Ratio of the normalized lepton $p_{l_\perp}$ (left) and dilepton $m_\perp$ (right) distributions, including the contribution from $\gamma\gamma$--initiated production (i.e. $\gamma\gamma$ + DY), to the pure DY case. In addition ratios of the pure DY with the $Z$ boson mass shifted by $10$ or $20$ MeV to the case with the default mass value are shown. The DY cross sections are calculated using fixed--order NLO QCD, while for the $\gamma\gamma$ case both the SF (with experimental uncertainties shown but generally not visible) and collinear \texttt{MMHT2015qed\_nnlo} predictions (with scale variation + PDF uncertainties) are given.}
\label{fig:MW}
\end{center}
\end{figure}

\begin{figure}
\begin{center}
\includegraphics[scale=0.64]{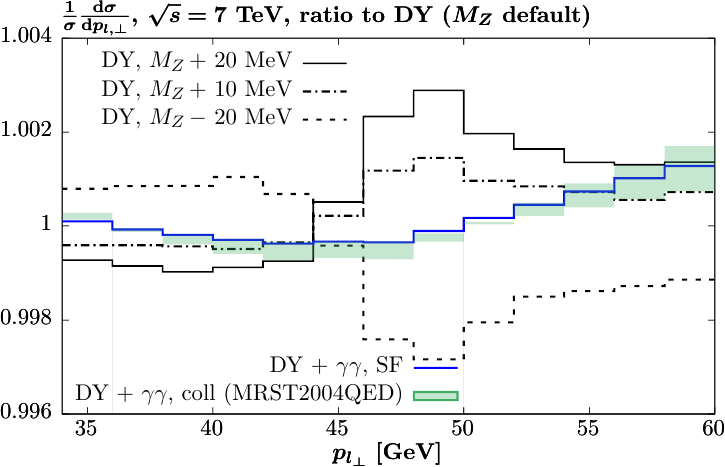}
\includegraphics[scale=0.64]{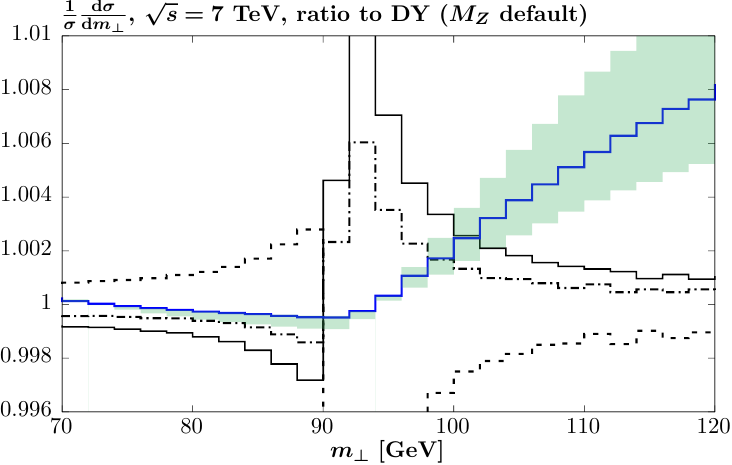}
\caption{As in Fig.~\ref{fig:MW}, but with collinear \texttt{MRST2004QED} predictions given. The PDF uncertainty corresponds to the difference between the constituent and current quark models, while scale variation uncertainties are omitted for clarity.}
\label{fig:MW1}
\end{center}
\end{figure}

As discussed in the previous section the contribution from $\gamma\gamma$--initiated production in the $Z$ peak region is small, though at low $p_\perp^{ll}$ as large as $\sim 4$ per mille. The precise quantitive impact of this contribution is non--trivial to evaluate, as it will depend on the details of the tuning performed by any given experiment, through which the effect will propagate.
Here, to give some idea of the relative size of effects, we will consider the relative contributions from the $\gamma\gamma$ channel to the $p_{l_\perp}$ and $Z$ boson $m_\perp$ distributions. Note the results are identical with respect to which lepton is considered to correspond to the `charged' one in defining these observables in this channel. This is shown in Fig.~\ref{fig:MW}, which in particular gives the ratio of the DY prediction including  $\gamma\gamma$--initiated production to the pure $Z/\gamma$ DY case. The basic idea in these comparisons is that any $\gamma\gamma$ contribution that is incorrectly tuned away in the comparison to the $Z/\gamma$ data could appear as a corresponding shift with respect to the description of the $W$ data. For example, if one did not subtract any PI component from the dilepton data, and one assumed that the underlying $l\/nu$ data corresponded to e.g. the PDG value of $M_W$, then the template at this value could still be shifted by the amount (and in the direction) shown in the figure.
We consider the normalized distributions to be the more relevant observables for such a comparison, although the conclusions are rather similar, though generally larger, if the absolute are instead taken. For the DY predictions we use \texttt{MCFM 6.8}~\cite{Campbell:2002tg,Campbell:2004ch} to provide fixed--order NLO QCD predictions. Given that the precise shapes of the distributions are sensitive to e.g. parton--shower effects, the following results should therefore be taken as a guide only.

To give an idea of the required precision, we in addition show the impact on the distributions of modifying the $Z$ boson mass by $10-20$ MeV away from the PDG value~\cite{PDG2019}; the corresponding modifications in the $W$ boson case follow these very closely after shifting the distributions to account for the different $W$ mass. Considering first the $p_{l_\perp}$ distribution in the left plot, we can see that impact of $\gamma\gamma$-initiated production on the normalized distribution is at the $\sim 1$ per mille level or below around the most relevant $M_Z/2$ region, while at higher values it is as large as $\sim 2$ per mille. It should be noted though that the higher $p_{l_\perp}$  region is likely to have a smaller impact on the corresponding tune, and in addition the underlying DY prediction will be more sensitive to effects beyond fixed-order NLO QCD. In any case, the effect is as we might expect rather small. However, we can see that while the effect of changing the input $Z$ mass by $10-20$ MeV is generally larger, the $\gamma\gamma$ contribution is not completely negligible in comparison. A similar conclusion is seen for the $m_\perp$ distribution in the right figure. Again the impact of the $\gamma\gamma$--initiated production is small, but  not negligible in comparison to the small impact of shifting the $Z$ mass, and arguably somewhat larger than in the $p_{l_\perp}$ case. In both plots the shape of the $\gamma\gamma$ induced modification does not follow that of the mass shifts particularly closely, as we might expect.

Now, as discussed above for the ATLAS measurement the PI component is already subtracted from the dilepton data~\cite{Aad:2014xaa} and thus this component will certainly not be fully absorbed into the tune. However, this subtraction is done using LO collinear theory, and with the outdated MRST2004QED set; both of these will lead to imprecision in the corresponding modelling of PI production. To assess the latter effect, we show in Fig.~\ref{fig:MW1} the same plots as before, but now with the LO collinear predictions made using this PDF set, and with $\mu_0=m_{ll}$. The PDF uncertainties alone, due to the difference between the constituent and current quark models, are shown for clarity; the scale variation uncertainty will as usual be present in addition. We can see that these are non--negligible, and as we would expect significantly larger than the SF uncertainties, though the results overlap within uncertainties in most regions. Now, in future analyses this outdated set will not be used, and hence a potentially more instructive comparison is as in Fig.~\ref{fig:MW}, i.e. with respect to the more precise and newer \texttt{MMHT2015qed\_nnlo} PDF result. However, here again we can see a clear difference between the SF and collinear LO predictions. We can in particular see that in many regions the SF prediction does not overlap with the LO collinear within scale variation bands of the latter prediction. This is not  surprising, given the sensitivity of these distributions to variables such as the dilepton $p_\perp^{ll}$, which is only accounted for in a non--trivial way by the SF prediction. As discussed in the previous section, one can of course improve the situation by going to higher orders in the collinear calculation, but the result of Fig.~\ref{fig:theorylowpt} (right) suggests that the contribution from elastic and low scale inelastic production may be particularly relevant at this level of precision, and the full kinematic dependence of these components can never be accounted for by a fixed--order collinear calculation. It is in particular worthy of note that in this comparison the difference between the LO collinear and SF predictions is sometimes comparable to the overall expected impact of PI production on these distributions.

In fact, in ATLAS analyses such as~\cite{Aad:2014xaa} the primary vertex is required to be reconstructed from at least three tracks. This will automatically exclude a significant fraction of PI events, in particular the majority of those due to elastic photon emission from the protons as well as a reasonable fraction of events with inelastic photon emission. A very useful feature of the 
SF calculation is that one readily isolate the contributions from elastic and inelastic photon emission and hence account for the impact of such a cut; elastic events will to first approximation always fail the vertex requirement, while for inelastic events this will depend on the kinematics of the proton dissociation system and its interface to an all--purpose MC generator for showering and hadronization of the system, all of which is accounted for in \texttt{SFGen}. In more detail however, one must account for the probability of no additional particle production from additional proton--proton interactions, independent of the hard scatter; that is, the so--called survival factor. This effect is not included in  \texttt{SFGen}, but these issues are discussed in detail in~\cite{Harland-Lang:2020veo}, with the \texttt{SuperChic} MC providing an implementation of the SF calculation of semi--exclusive PI lepton pair production relevant for events in the presence of the vertex requirement imposed by ATLAS.

 We end this section by again noting that the above results can only serve as a first look at the possible impact of these effects, in particular given the connection between the tuning to dilepton data and the $M_W$ determination is not completely direct, and will as mentioned above depend on the details of the tuning and the precise validation performed, as well as the fact that we only use fixed order NLO QCD predictions here.  An alternative, and potentially informative approach would be to consider the effect of modifying the $W$ boson $p_\perp$ distribution according to  Fig.~\ref{fig:theorylowpt} (right) directly on the extracted $W$ mass; such modifications at low $p_\perp$ have been examined in e.g.~\cite{Bacchetta:2018lna} in the distinct context of intrinsic parton $k_\perp$. In any case, our results certainly verify that $\gamma\gamma$-initiated production is potentially relevant at the level of precision being aimed for, and in future experimental analyses the SF approach could provide an invaluable tool to account for this.

\subsection{Triple--differential Drell Yan: photon--initiated contribution}\label{sec:ll3d}

\begin{figure}[t]
\begin{center}
\includegraphics[scale=0.5]{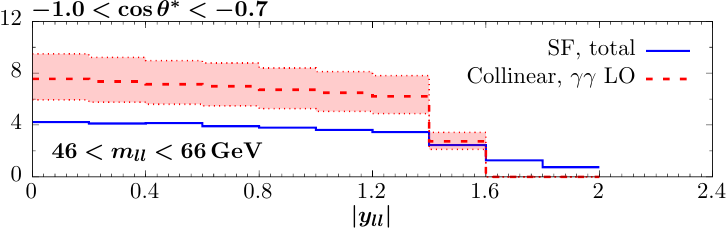}
\includegraphics[scale=0.5]{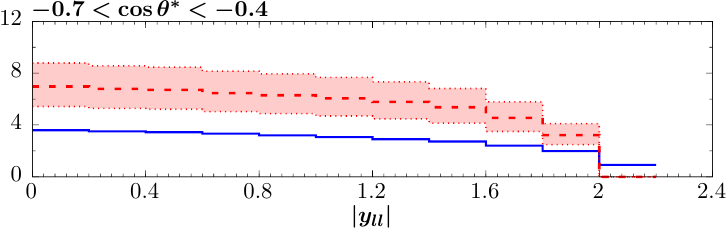}
\includegraphics[scale=0.5]{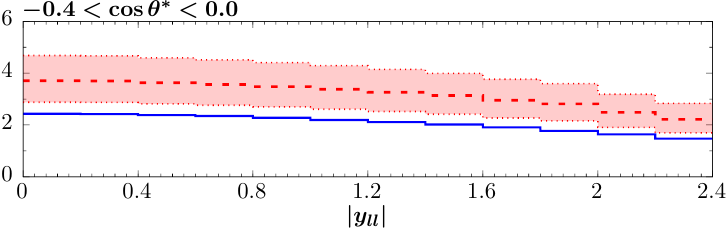}
\includegraphics[scale=0.5]{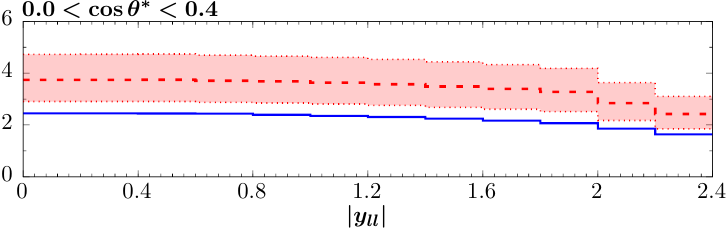}
\includegraphics[scale=0.5]{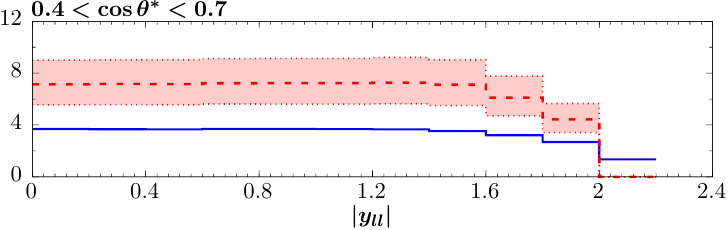}
\includegraphics[scale=0.5]{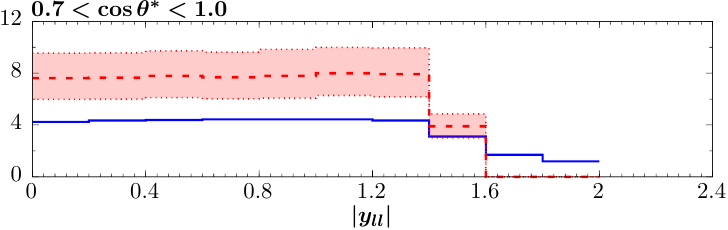}
\caption{Percentage contribution from PI production to the pure DY cross section. The lepton pair $y_{ll}$ distribution is shown, in different bins of $\cos\theta^*$, and  for $46 < m_{ll} < 66$ GeV. The DY result is calculated at NNLO in QCD, while the PI cross section corresponds to the SF and collinear LO calculations. In the SF case the error band due to the experimental uncertainty on the structure functions is shown (note in many regions this is comparable to the line width of the central value), while in the collinear case  that due to scale variation and PDF uncertainties added in quadrature is given, with the former being strongly dominant. The event selection corresponds to the ATLAS~\cite{Aaboud:2017ffb} central lepton measurement. In all figures in this section, we only show values for those bins where a non--zero measurement is recorded by ATLAS.}
\label{fig:Z3D_m46_66cc}
\end{center}
\end{figure}

\begin{figure}[t]
\begin{center}
\includegraphics[scale=0.5]{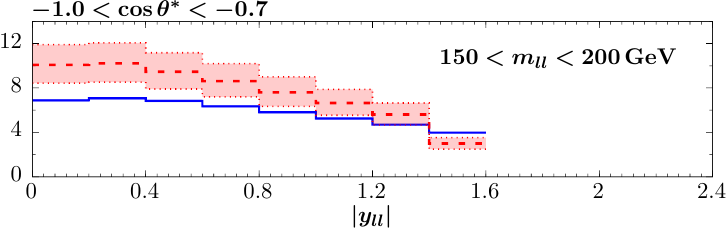}
\includegraphics[scale=0.5]{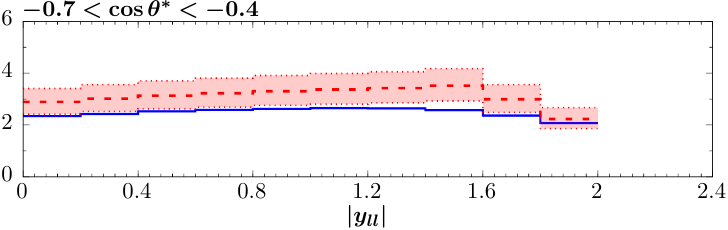}
\includegraphics[scale=0.5]{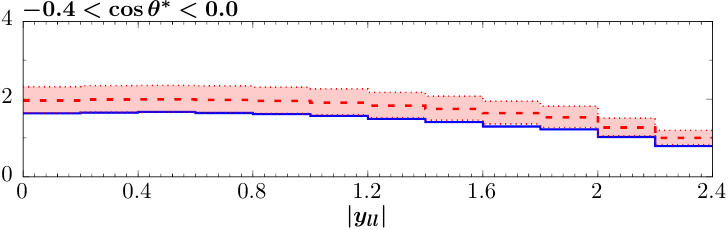}
\includegraphics[scale=0.5]{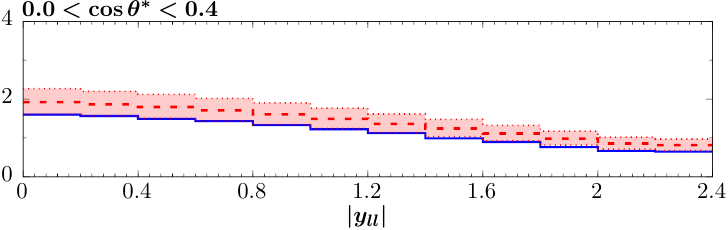}
\includegraphics[scale=0.5]{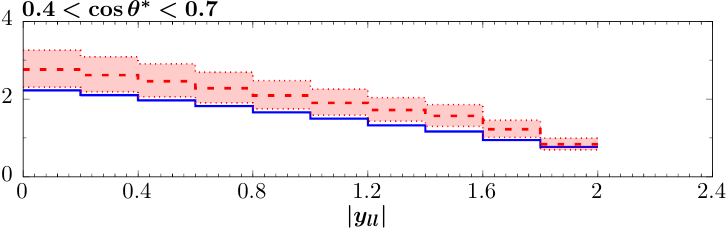}
\includegraphics[scale=0.5]{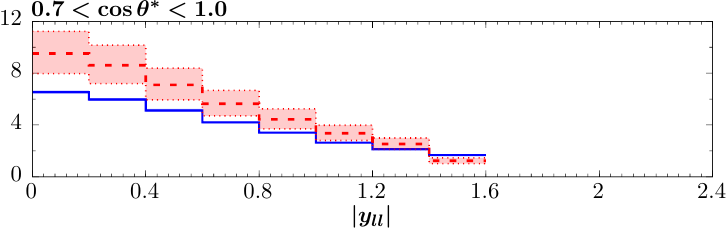}
\caption{As in Fig.~\ref{fig:Z3D_m46_66cc}, but for $150 < m_{ll} < 200$ GeV. }
\label{fig:Z3D_m150_200cc}
\end{center}
\end{figure}

High precision triple--differential measurements of lepton pair production have been reported at 8 TeV by both ATLAS~\cite{Aaboud:2017ffb} and CMS~\cite{Sirunyan:2018swq}, with important implications both for PDF constraints and the determination of the weak mixing angle, $\sin^2 \theta_W$. There are in particular given in terms of the dilepton rapidity, $y_{ll}$, invariant mass $m_{ll}$, and $\cos\theta^*$ (and/or the forward--backward asymmetry constructed from this), where $\theta^*$ is the lepton decay angle in the Collins--Soper frame~\cite{Collins:1977iv}. We can expect PI production to give an important contribution in certain regions of phase space, and therefore for this to have an impact on any determination of $\sin^2 \theta_W$ as well as PDF constraints from the data. While, as discussed in~\cite{Sirunyan:2018swq}, the former is mostly sensitive to the $Z$ peak region, where the PI contribution is smallest, sensitivity is not completely confined to this, and conversely the largest sensitivity to PDFs comes from the off peak regions. Moreover the approach applied in the CMS analysis is for in--situ constraints on the PDFs to be performed across the considered $m_{ll}$ region in order to maximise the precision of the measurement of $\sin^2 \theta_W$. Thus even if the relative contribution from PI production is quite small on the $Z$ peak, it could still bias such a procedure if not accounted for correctly.

In this section we therefore show some representative results for the PI contribution to triple differential dilepton production, considering the ATLAS 8 TeV event selection for concreteness. We will in particular present results for the percentage contribution from PI production to the NNLO QCD DY prediction, as calculated using \texttt{NNLOJET}~\cite{Bizon:2018foh}. We leave a comparison to data to future work, as this can most meaningfully be done in the context of a global PDF fit, accounting for all sources of correlated error in the data. In all cases, we only show the pure $\gamma\gamma$ channel for clarity, but have checked that the contribution from $Z$ bosons and the mixed $\gamma/Z + q$ channel is small.  We only show values for those bins where a non--zero measurement is recorded by ATLAS, in order to avoid regions of phase space where the cross section is very highly suppressed and hence the ratio values may be somewhat misleading. In addition, stable NNLO QCD predictions for some of these bins were not always available.

We first note that in general the forward--backward asymmetry
\be
A_{FP} = \frac{\sigma_F- \sigma_B}{\sigma_F + \sigma_B}\;,
\ee
 in PI production is of course zero; here $\sigma_F$ and $\sigma_B$ are the cross sections in the forward ($\cos\theta^*>0$) and backward ($\cos\theta^*<0$) hemispheres, respectively. However, the contribution from this to the overall dilepton sample is not, as it will increase the denominator in the above, while leaving the numerator unaffected. In other words, it will tend to wash out the asymmetry. More generally, it will affect the fully differential cross section in $\cos\theta^*$ in a non--trivial way.
 
 \begin{figure}
\begin{center}
\includegraphics[scale=0.5]{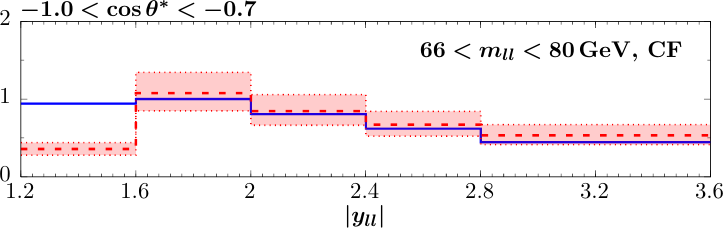}
\includegraphics[scale=0.5]{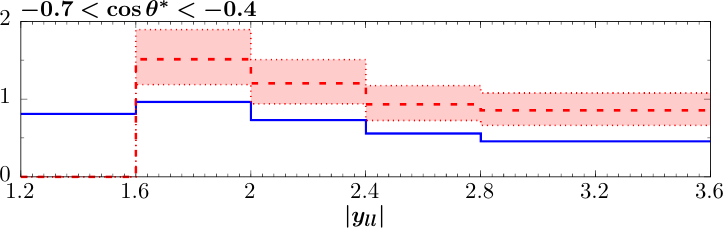}
\includegraphics[scale=0.5]{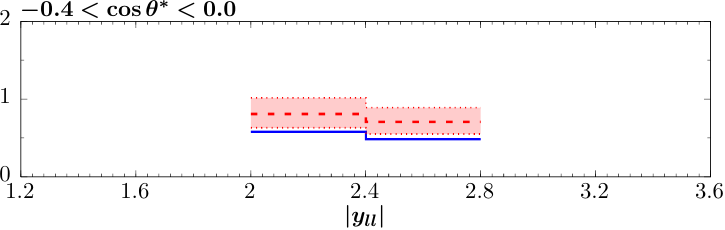}
\includegraphics[scale=0.5]{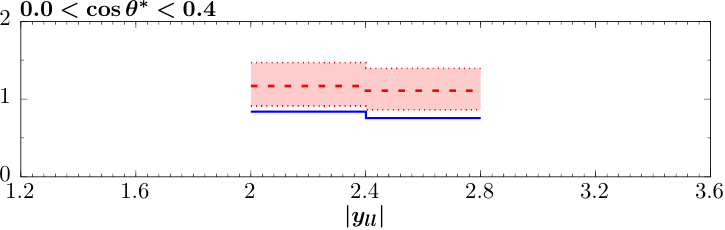}
\includegraphics[scale=0.5]{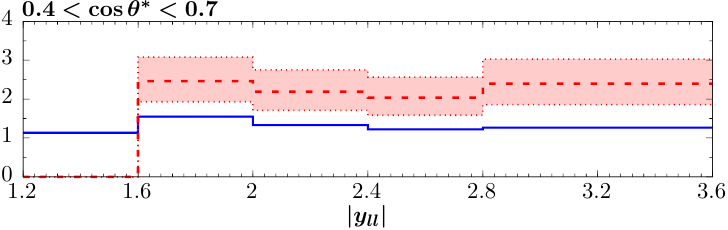}
\includegraphics[scale=0.5]{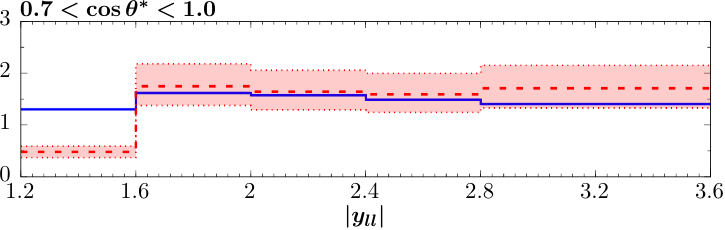}
\caption{As in Fig.~\ref{fig:Z3D_m46_66cc}, but for $66 < m_{ll} < 80$ GeV and the  ATLAS~\cite{Aaboud:2017ffb} central--forward lepton measurement.}
\label{fig:Z3D_m66_80cf}
\end{center}
\end{figure}

\begin{figure}
\begin{center}
\includegraphics[scale=0.5]{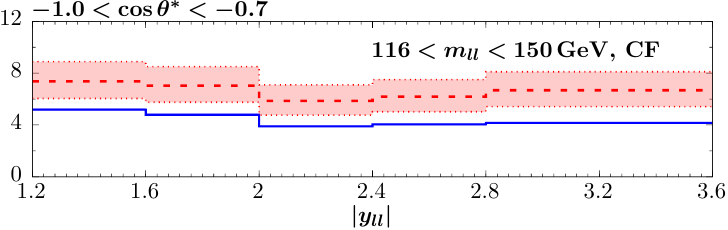}
\includegraphics[scale=0.5]{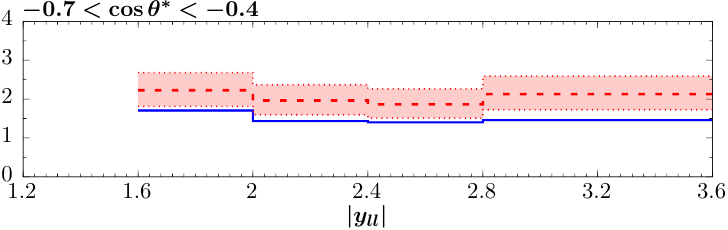}
\includegraphics[scale=0.5]{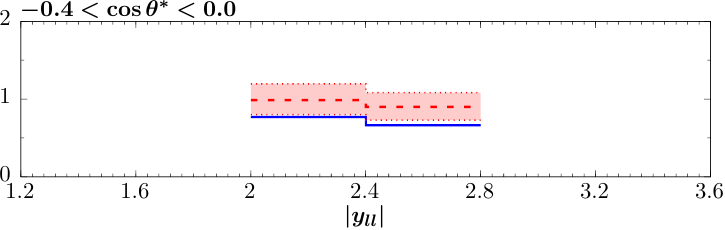}
\includegraphics[scale=0.5]{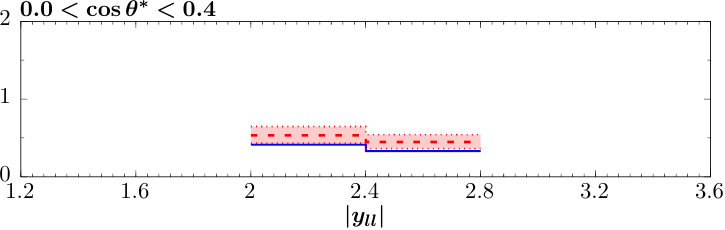}
\includegraphics[scale=0.5]{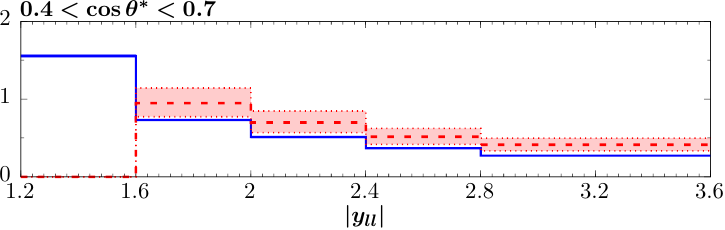}
\includegraphics[scale=0.5]{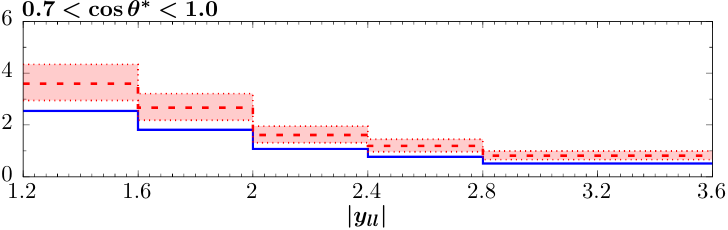}
\caption{As in Fig.~\ref{fig:Z3D_m66_80cf}, but for $116 < m_{ll} < 150$ GeV.}
\label{fig:Z3D_m116_150cf}
\end{center}
\end{figure}

We will focus on those mass bins where the PI contribution is largest, namely at the lower and upper ends of the considered range, but will comment on the contribution in other bins at the end of the section. We begin by consider the central lepton (muon and electron combined) measurement, that is with both leptons required to lie in the $|\eta_l| < 2.4$ region, with $p_{l_\perp} > 20$ GeV. Results for the $46 < m_{ll} < 66$ GeV and  $150 < m_{ll} < 200$ GeV mass bins are shown in Figs.~\ref{fig:Z3D_m46_66cc} and \ref{fig:Z3D_m150_200cc}, respectively. Again, though the PI contribution is symmetric in the $\cos\theta^*>0$ and $\cos\theta^*<0$ regions, the DY contribution is not, and hence these ratio plots are not symmetric with respect to this. We can see that the PI contribution ranges from $\sim 1 -6\%$ in all bins, and is therefore small but certainly not negligible. In more detail, there is a clear tendency for the  PI contribution to be larger at low rapidities and forward  $\cos\theta^*$ values; in the latter case this is driven by the $t$--channel nature of the PI process. Thus, as expected this will modify the corresponding $\cos\theta^*$ distribution and asymmetry $A_{FB}$. 

We in addition show the  LO collinear result, with corresponding scale variation uncertainties. Interestingly, there is a distinct tendency for the collinear prediction to lie above the SF, as seen in Fig.~\ref{fig:mDY} already. However, here the effect is larger and the LO result lies above the SF outside the scale variation uncertainty bands in most bins. We note that the increased size of this difference  results from the event selection being different in comparison to Fig.~\ref{fig:mDY}, for which no cut on the $p_\perp$ of the lepton is placed. Indeed, given the LO collinear prediction corresponds in essence to a cross section integrated over the photon virtualities $Q^2$, the less inclusive the observable becomes, the larger we might expect any disagreement with the SF result to be, and this is precisely what we see. In some of the higher $y_{ll}$ bins there is no phase space for production with zero dilepton $p_{\perp}^{ll}$, and hence the collinear prediction vanishes entirely. This is in contrast to the SF result, though these are in rather suppressed regions, where moreover the QCD prediction may be less reliable. 

We next consider results for the central--forward electron selection, that is with the central electron required to have $p_{l_\perp} > 25$ GeV and $|\eta_l| < 2.4$, while the forward electron is required to have $p_{l_\perp} > 20$ GeV and $2.5<|\eta_l| < 4.9$. Results for  $66 < m_{ll} < 80$ GeV and  $116 < m_{ll} < 150$ GeV are shown in Figs.~\ref{fig:Z3D_m66_80cf} and \ref{fig:Z3D_m116_150cf}, respectively. Although the precise shapes of the distributions is different, the broad picture is very similar to the central channel. Namely, the PI contribution ranges from $\sim 1 -6\%$, with some tendency to decrease with $y_{ll}$, though this effect is certainly milder here. The LO collinear predictions again have large scale variation uncertainties and lie above the SF predictions, often outside of the uncertainty band. As before, in certain  bins the LO collinear result can be zero due to kinematics, while not being the case in the SF calculation. 

Finally, we comment on the mass bins not shown here, for which the PI contribution is generally smaller. For the central selection, the off--peak bins of  $66 < m_{ll} < 80$ GeV and  $102 < m_{ll} < 116$ GeV have PI contributions of $\sim 1-2\%$ and $\sim 1-4\%$, respectively. For the two on--peak bins, $80 < m_{ll} < 91$ GeV and $91 < m_{ll} < 102$ GeV, the contribution is at the $\lesssim 1$ per mille level. For the central--forward selection, the off--peak bin  $102 < m_{ll} < 116$ GeV has a PI contribution of $\sim 1\%$ or less, while for the on--peak bins this is similar to the central case.

\section{Lepton--lepton scattering and the lepton PDF}\label{sec:llscat}

\begin{figure}
\begin{center}
\subfigure[]{\includegraphics[scale=0.8]{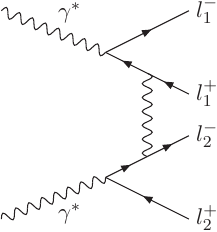}}\qquad\qquad
\subfigure[]{\includegraphics[scale=0.8]{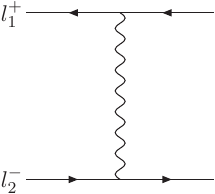}}
\caption{Representative diagram for lepton--lepton scattering in (a) the SF formalism and (b) the corresponding diagram in the collinear lepton PDF formalism.}
\label{fig:FDlep}
\end{center}
\end{figure}

In this section we consider an application of the SF formalism to a different class of process to that considered above. Namely, one where a lepton produced in  a $\gamma^* \to l^+ l^-$ splitting can be viewed as an initial--state parton in the production process. This idea has recently been considered in~\cite{Buonocore:2020nai}, where it shown that  by integrating over the initial--state photon kinematics one can recast the process in terms of a collinear lepton PDF within the proton. As with the original analysis of~\cite{Manohar:2016nzj,Manohar:2017eqh} for the photon PDF, they demonstrate how this can be defined consistently such that higher order contributions, i.e. those with an explicit $\gamma \to l^+ l^-$ splitting, where the initial--state photon is treated in the collinear LO approach, can be included in the $\overline{{\rm MS}}$ scheme. However, given our previous results in the context of PI production in the SF calculation, we may expect to arrive at a rather precise evaluation of this process by instead using the SF calculation directly. 

To demonstrate this, we reconsider the example of lepton--lepton scattering taken in~\cite{Buonocore:2020nai}. As discussed there, the observation of  an isolated and back--to--back same--sign lepton pair of the same flavour, or a lepton pair of differing flavours and arbitrary signs is an important test of flavour violating interactions in various BSM scenarios, see~\cite{ATLAS:2014kca,Aaboud:2017qph,Sirunyan:2018xiv,Sirunyan:2020ztc}. A representative diagram for the lepton--lepton scattering subprocess within the SF approach is shown in Fig.~\ref{fig:FDlep} (a), while the corresponding LO collinear diagram is given in Fig.~\ref{fig:FDlep} (b). The calculation of the SF cross section then requires a relatively straightforward application of \eqref{eq:sighh}; we will drop the $Z$ and mixed $Z/\gamma + q$ contributions, which are both negligible. In a little more detail, some care is needed in performing the phase space integration, as there are two independent logarithmic enhancements in the region of collinear $\gamma \to l^+ l^-$ emission from both initial--state legs, which while regulated by the non--zero photon $Q^2$ as well as the lepton mass, and so perfectly finite, can destabilise the numerical integration. The solution is simply to express the phase space  integration directly in terms of the two leptons we do not require to be observed in the detector, and hence which can/will be emitted favourably in a collinear direction to the initial-state photon. Their angular integration is then alway defined  with respect to the photon direction, which for off--shell photons is not in general along the $z$ axis. Once this is done, standard adaptive integration tools and/or a suitable change of variables can stabilise this logarithmic enhancement.

We consider precisely the same observables and event selection as in~\cite{Buonocore:2020nai}, in order to perform a direct comparison. In particular, the signal leptons are required to have $p_{l_\perp}> 20$ GeV and $|\eta_l|<2.4$ and results at 13 and 27 TeV are shown. The cross section results are shown in Table~\ref{tab:lepscatt} for both the SF calculation and the LO collinear case, as given in~\cite{Buonocore:2020nai}. Errors due to the experimental uncertainty on the structure function are given in the SF case, and factorization scale variation in the collinear case. We can see that the LO collinear predictions  carry significant errors due to this, ranging from $\sim 30-100\%$ of the cross section, which are absent in the SF case. The errors on the latter are as expected very small, at the percent level. 
The SF results lie somewhat below the central LO collinear predictions, but within scale variation uncertainties in all cases. 
 In terms of the scaling with the mass of the leptons in the final state, these are closely matched in the SF and LO calculations, e.g. the ratio of the $e^+ \mu^-$ to $e^+ \tau^-$ cross section is rather similar in the two cases. 
 As pointed out in~\cite{Buonocore:2020nai}, a NLO calculation can certainly be performed, which will result in smaller scale variations in the collinear case and, we would expect, a somewhat closer agreement with the SF prediction.

\renewcommand{\arraystretch}{1.5}
\begin{table}
\begin{center}
\begin{tabular}{|c|c|c|c|c|c|c|}
\hline
 & $e^+ \mu^-$  &$e^+ \tau^-$  &$\mu^+ \tau^-$  & $e^+e^+$& $\mu^+ \mu^+$  &$\tau^+ \tau^+$  \\
\hline 
13 TeV, Ref.~\cite{Buonocore:2020nai} & $0.29^{+0.13}_{-0.10}$ & $0.18^{+0.11}_{-0.08}$ & $0.16^{+0.10}_{-0.07}$ & $0.24^{+0.10}_{-0.08}$ & $0.19^{+0.09}_{-0.07}$ & $0.08^{+0.06}_{-0.04}$ \\ 
\hline
13 TeV, this work & $0.254^{+0.004}_{-0.004}$ & $0.158^{+0.002}_{-0.002}$ & $0.140^{+0.002}_{-0.002}$ & $0.181^{+0.003}_{-0.003}$ & $0.140^{+0.002}_{-0.002}$  & $0.054^{+0.001}_{-0.001}$  \\ 
\hline
27 TeV, Ref.~\cite{Buonocore:2020nai}& $0.53^{+0.25}_{-0.18}$ & $0.34^{+0.21}_{-0.15}$ & $0.30^{+0.19}_{-0.14}$ & $0.440^{+0.19}_{-0.14}$ & $0.34^{+0.16}_{-0.12}$ & $0.14^{+0.12}_{-0.07}$ \\ 
\hline 
27 TeV, this work & $0.482^{+0.008}_{-0.008}$ & $0.304^{+0.004}_{-0.004}$ & $0.268^{+0.004}_{-0.004}$ & $0.347^{+0.002}_{-0.002}$ & $0.259^{+0.004}_{-0.004}$ & $0.104^{+0.002}_{-0.002}$ \\ 
\hline
\end{tabular}
\end{center}
\caption{Cross sections (in fb) for same--sign same flavour and opposite--sign different--flavour lepton pair production in the SM. Results are shown both for the SF and the collinear LO approach, with the latter given in terms of a lepton PDF in the proton. For the SF values, errors due to the experimental uncertainty on the structure functions are given, while in the collinear result factorization scale variations are given. Note the collinear results correspond to those presented in~\cite{Buonocore:2020nai}.}  \label{tab:lepscatt}
\end{table}

\begin{table}
\begin{center}
\begin{tabular}{|c|c|c|c|c|c|c|}
\hline
 & $e^+ \mu^-$  &$e^+ \tau^-$  &$\mu^+ \tau^-$  & $e^+e^+$& $\mu^+ \mu^+$  &$\tau^+ \tau^+$  \\
\hline
13 TeV & 0.32 & 0.17 & 0.16 & 0.37 & 0.30  &0.10 \\ 
\hline
27 TeV & 0.31 & 0.16 & 0.15 & 0.35 & 0.28 & 0.10 \\ 
\hline
\end{tabular}
\end{center}
\caption{Acceptance for same--sign same flavour and opposite--sign different--flavour lepton pairs in the SM, after applying the transverse momentum balance and acoplanarity cuts described in the text. Results given for the SF calculation only; in the LO collinear case the acceptance is by definition 1.0.}  \label{tab:lepscattcut}
\end{table}

\begin{table}
\begin{center}
\begin{tabular}{|c|c|c|c|c|c|c|}
\hline
& $e^+ \mu^-$  (el.) &$e^+ \mu^-$  (sd)  &$e^+ \mu^-$  (dd) &$\tau^+ \tau^+$ (el.) &$\tau^+ \tau^+$ (sd) &$\tau^+ \tau^+$  (dd) \\
\hline
$\sigma$ [fb] & 0.0832 & 0.121 & 0.0460 & 0.0102 & 0.0262  &0.0170 \\ 
\hline
Acceptance & 0.50 & 0.27 & 0.15 & 0.15 & 0.091 & 0.075 \\ 
\hline
\end{tabular}
\end{center}
\caption{Cross sections  (in fb) and acceptance corrections (as defined in Table~\ref{tab:lepscattcut} and the text) for same--sign same flavour and opposite--sign different--flavour lepton pairs in the SM, at 13 TeV. Results within the SF approach and for the differing elastic (el.), single dissociation (sd) and double dissociation (dd) cases are given.}  \label{tab:lepscattbd}
\end{table}

Next, we consider the impact of the cuts
\be\label{eq:b2b}
\frac{|\vec{p}_{l_{1\perp}}+\vec{p}_{l_{2\perp}}|}{{\rm max}(p_{l_{1\perp}},p_{l_{2,\perp}})} < 0.1\;,\qquad |\Delta \phi_{l_1 l_2}- \pi | < 0.1\;,
\ee
which as described in~\cite{Buonocore:2020nai} can strongly suppress the dominant SM contribution from same--sign $W$ pair plus dijet production. While these leave the LO collinear prediction  trivially unchanged, by applying the SF calculation we can get a precise evaluation of their impact, shown in Table~\ref{tab:lepscattcut}. No errors from the experimental uncertainty on the structure functions are given, as these will be tiny. We can see that impact of the cuts is expected to be significant, decreasing the cross section by a factor of $\sim 3-10$, depending on the process. This is not necessarily surprising, as the pure back--to--back configuration of the LO collinear prediction will always be modified by the non--zero photon $Q^2$ in the full SF calculation as well the non--zero opening angle in the $\gamma^* \to l^+ l^-$ transition. We find that the impact of the cuts is larger for the heavier leptons, as one might expect from the fact that the collinear enhancement in the $\gamma^* \to l^+ l^-$ transition is less strong in this case. There is a very mild decrease in the acceptance at the larger $\sqrt{s}=27$ TeV value, perhaps due to the increased phase space in photon $Q^2$ at the lower photon $x$ value this corresponds to. 

To analyse these effects in a little more detail, in Table~\ref{tab:lepscattbd} we show the acceptance (and corresponding cross sections values) for $e^+ \mu^-$ and $\tau^+ \tau^+$ production, but now breaking things down in terms of whether the initial--state photons are emitted elastically or not. We can see that even for purely elastic emission, where the average photon $Q^2$ is rather low, only 50\% of events survive the cuts. As we expect, when one proton  dissociates (`sd') this acceptance reduces, and when both protons dissociate (`dd') this reduces further still. The combination of these channels, with their respective acceptances, gives the corresponding values in Table~\ref{tab:lepscattcut}. 

The above results again highlight a useful feature of the SF calculation, namely that one can readily isolate the contribution from elastic and inelastic photon emission. Indeed, this can provide a useful handle in searches for processes like lepton pair production discussed above. In particular, one can use the fact that additional hadronic activity associated with the event is strongly suppressed for the SM lepton--lepton scattering process, whereas it might not be for a BSM signal, or for other SM contributions to the final state. However, a precise account of this requires that we include the effect of multi--parton interactions, i.e. the so--called soft survival factor, if we require no additional hadronic activity, as well as the fact that the proton dissociation products may be present in the detector for inelastic photon emission. These issues are discussed in detail in~\cite{Harland-Lang:2020veo}, where an approach that accounts for these effects in (opposite sign, same flavour) lepton pair production is presented and implemented in the \texttt{SuperChic 4} MC. Such an approach could straightforwardly be applied to the current scenario, but we leave that for future work.

We note that in the above calculations there is a further possible contribution from $\gamma \gamma \to l^+_1 l^-_1$ production, with a subsequent $\gamma \to l^+_2 l^-_2$ emission from the lepton type 1 line. For example, for $\mu^+ \mu^+$ production we would have $l_1=l_2=\mu$. Considering this example in more detail, the $\mu^+$ from the original $\gamma \gamma \to l^+_1 l^-_1$ would have to produced at sufficiently high $p_\perp$ to pass the experimental cuts, and only the $\mu^-$ emission would receive a collinear enhancement, being dominantly emitted at low $p_\perp$. Then, the dominant contribution from the $\gamma \to l^+_2 l^-_2$ would be due to collinear emission. To deal with this contribution one would therefore need to impose suitable isolation requirements on the signal leptons. This would remove this collinear enhancement but potentially leave this as a genuine NLO correction. However, once the back--to--back requirements \eqref{eq:b2b} are imposed such a contribution should be very small. In particular, and even absent any lepton isolation conditions, these will only be passed if the lepton $l^-_2$ ($=\mu^-$) is at very low $p_\perp \lesssim O({\rm GeV})$ in order that the two signal leptons are back--to--back. This is a rather stringent requirement that should reduce such a contribution significantly below the level one would naively expect from a NLO contribution. We therefore do not include this contribution in the calculation, though it could certainly be included within the SF approach.

Finally, we note that the experimental uncertainties on the structure functions are not the only source of uncertainty in the SF prediction. As in the case of lepton pair production, we should also in general include the uncertainty from non--factorizeable corrections, as well as higher order uncertainties in the pQCD calculation of the structure functions, which as before will enter at the percent level. We in addition should consider the impact of higher order QED (as well as in principle EW/QCD) corrections to the underlying $\gamma \gamma$ to four lepton process. This raises a possible shortcoming of the SF approach, namely that  photon emission from the leptons initiating the lepton--lepton scattering process receives a collinear logarithmic enhancement. This will be resummed via the DGLAP evolution of the lepton PDFs in the collinear calculation, but this would not happen in the SF calculation outlined above. 

Such a correction is discussed in detail in~\cite{Buonocore:2020nai}, Appendix D, where the impact of allowing a single photon emission on the derived lepton PDFs is considered. Even up to rather large scales, $\mu_F \sim$ 1 TeV, the effect is small, at the percent level or less apart from at rather high $x$ values. This level of difference, driven by the small size of the QED coupling, is not indicative of a strong degree of logarithmic enhancement and with it the necessity of performing a resummation. Though the impact is larger at high $x$, here the experimental uncertainty on the structure functions is larger as well, so the phenomenological impact (and indeed relevance) appears to be relatively limited. Moreover, one can see in~\cite{Buonocore:2020nai} that there is a $\sim$ percent level uncertainty in precisely how one assigns such a logarithmically enhanced contribution to the lepton PDF, which is smaller than, but not negligible in comparison to, the overall size of the effect. It is certainly not clear that this is a smaller difference than that between only including one photon emission explicitly, say, in the SF approach, and resumming these  contributions, as is done in the DGLAP evolution of the lepton PDFs.

Indeed, one can of course include an additional photon emission explicitly in the SF calculation if the desired level of precision requires it, and in principle one could even resum the effect of such emissions directly in the SF calculation. This would arguably  be more cumbersome than achieving the desired resummation in the collinear calculation, i.e. via the DGLAP evolution of the lepton PDFs. However, as the impact of such effect is in any case rather small, and certainly significantly smaller than the factorization scale variations in the LO collinear result, and quite possibly the NLO one, it is almost certainly a question of relatively marginal phenomenological relevance for LHC physics. This is particularly true for the comparisons presented above, where the relevant scale of the hard process is of order $p_{l_\perp}^{\rm cut}= 20$ GeV, which is rather low.

\section{\texttt{SFGen} Monte--Carlo: availability}\label{sec:SFgen}

The \texttt{SFGen} MC provides a publicly available tool for lepton pair production and lepton--lepton scattering within the SF approach, including initial--state $Z$ and mixed $Z/\gamma +q$ contributions. Arbitrary distributions and unweighted events can be generated, and errors due to the experimental uncertainty on the structure function (the equivalent of PDF uncertainties in the photon PDF framework) can be calculated on--the--fly. Unweighted events an be interfaced to \texttt{Pythia} for showering/hadronization of the proton dissociation system, following the approach discussed in~\cite{Harland-Lang:2020veo}. The LO collinear PI predictions can also be calculated for comparison, for arbitrary factorization scales. For the $Q^2>1\,{\rm GeV}^2$ continuum contribution, the code makes use of structure function grids in the \texttt{LHAPDF}~\cite{Buckley:2014ana} format, as calculated using \texttt{APFEL}~\cite{Bertone:2013vaa}, and first used in~\cite{Bertone:2018dse}. These structure function grids are provided with the MC, and are calculated using the approach discussed in Section~\ref{sec:gamzq}, namely with  \texttt{MMHT2015qed\_nnlo} PDFs~\cite{Harland-Lang:2019pla} in the high $Q^2$ region. The code and user manual is available at:
\\
\\
\url{http://projects.hepforge.org/SFGen}

\section{Summary and Outlook}\label{sec:conc}

The LHC and its high luminosity upgrade will provide a significant opportunity for stress testing the SM at an unprecedented level of precision. Events with leptons in the final state play a particularly important role in this, enabling for example the determination of the weak mixing angle, $\sin^2 \theta_W$, and the $W$ boson mass, $M_W$, as well as providing constraints on PDFs. The availability of precise theoretical calculations for SM processes is essential in achieving this, and a key element of this is the contribution from photons in the initial state, that is so--called photon--initiated (PI) production.

The study of~\cite{Harland-Lang:2019eai} showed for the first time how the PI production cross section of lepton pairs, away from the $Z$ peak, could be straightforwardly calculated with percent--level precision, by applying the structure function (SF) approach. In this paper, we have explored in more detail the phenomenological implications of this approach for LHC processes with leptons in the final state. We have calculated the PI contribution to multi--differential dilepton production, which is relevant for determinations of $\sin^2\theta_W$ as well as PDF constraints, and evaluated the potential impact  on $W$ mass determinations, through the tuning that is performed to dilepton events. In both cases, a precise account of PI production is found to be highly relevant, in particular given the level of precision being aimed for at the LHC. We have shown how the direct SF calculation automatically provides this. A key element in evaluating this contribution is control over all phase space regions, in particular at low dilepton $p_\perp^{ll}$, which cannot be accessed via a purely collinear calculation in terms of a photon PDF, and the SF calculation is therefore uniquely positioned to evaluate.

We have presented a detailed comparison of the SF result to the collinear calculation at the highest available order, namely NLO. We find that broadly an improved agreement with the SF result can be achieved with respect to the LO calculation, with correspondingly smaller scale variation uncertainties. However, these uncertainties (which are absent in the SF result) remain significantly larger than the percent--level PDF uncertainties, and related uncertainty due the experimental determination of the structure functions in the SF result, in many regions. Moreover, the agreement between the NLO prediction and the SF is not always within these scale variation bands, in particular for certain choices of scale. These issues are bypassed for pure $\gamma\gamma$ PI process, i.e. due to $t$--channel lepton exchange, which dominates away from the $Z$ peak, in the SF approach.

We have also extended the previous calculation to include contributions from $Z$ bosons in the initial state and  mixed $Z/\gamma +q$ contributions. In addition, the PI production of same--sign same--flavour, or different--flavours arbitrary--sign lepton pairs is calculated. We have found that the SF approach can indeed provide high precision predictions for this process, and the result of applying cuts in order to isolate back--to--back lepton topologies is readily evaluated. 

On the other hand, while the benefit of applying the SF approach is clear for process where the final state of interest ($l^+ l^-$, $W^+ W^-$...) is directly produced by the $\gamma\gamma$ initial state, this is not as clear for the mixed $\gamma + q$ contributions. Here, one must deal with the collinear enhancement of the $\gamma \to q\overline{q}$ splitting, and indeed further QCD splittings at higher orders, and at this level of precision include QED DGLAP evolution of the quark/antiquark PDFs. This certainly requires one introduce a photon PDF within the LUXqed approach. 
 Nonetheless, one may systematically account for this while still working within the SF approach. We in particular discuss the complication that arises due to double counting with the PI contribution that is already effectively included via the QED DGLAP evolution of the quark PDFs, and provide a simple procedure for removing this.  This however ties the level of precision for the SF calculation to that in the corresponding QED DGLAP evolution. It would  interesting in the future to examine the extent to which the two calculations agree or differ for these processes.

Finally, we have developed the  \texttt{SFGen} MC, a publicly available tool for  lepton pair and lepton--lepton scattering within the SF approach, including initial--state $Z$ and mixed $Z/\gamma +q$ contributions. Arbitrary distributions and unweighted events can be generated, and errors due to the experimental uncertainty on the structure function (the equivalent of PDF uncertainties in the photon PDF framework) can be calculated on--the--fly.

In summary, the PI mechanism provides a moderate but important contribution to lepton pair production. Given the high precision being aimed for at the LHC, a precise account of it is therefore mandatory. We have shown that the SF approach provides a way to achieve this, and demonstrated how this can be extended to provide high precision predictions for other processes involving leptons, such as lepton--lepton scattering. To enable the application of this approach to calculations of such processes, we have provided a publicly available MC tool for use by the community.
 
\section*{Acknowledgments.}

I thank Valery Khoze, Misha Ryskin and Robert Thorne for useful discussions and for passing on comments about the manuscript. I thank Marco Zaro for guidance with \texttt{MadGraph5\_aMC@NLO}. I thank Valerio Bertone for guidance in the use of \texttt{APFEL} and Alexander Huss for providing NNLO and NNLO+N${}^3$LL QCD predictions for the Drell Yan production. I thank Maarten Boonekamp for useful clarifications about the treatment of PI production in ATLAS analyses. I thank Giulia Zanderighi and collaborators for helping to identify a missing factor of two in the calculation of the different flavour lepton scattering cross sections, and for pointing out the possible contribution from $\gamma \gamma \to l_1^+ l_1^- (+ \gamma \to l^+_2 l^-_2)$.  I thank the Science and Technology Facilities Council (STFC) for support via grant award ST/L000377/1.

\bibliography{references}{}

\begin{thebibliography}{10}

\bibitem{Aaij:2015gna}
LHCb, R.~Aaij {\em et~al.},
\newblock JHEP {\bf 08}, 039 (2015), 1505.07024.

\bibitem{Aaij:2015zlq}
LHCb, R.~Aaij {\em et~al.},
\newblock JHEP {\bf 01}, 155 (2016), 1511.08039.

\bibitem{Aaij:2015vua}
LHCb, R.~Aaij {\em et~al.},
\newblock JHEP {\bf 05}, 109 (2015), 1503.00963.

\bibitem{Khachatryan:2016pev}
CMS, V.~Khachatryan {\em et~al.},
\newblock Eur. Phys. J. C {\bf 76}, 469 (2016), 1603.01803.

\bibitem{Aaboud:2016btc}
ATLAS, M.~Aaboud {\em et~al.},
\newblock Eur. Phys. J. C {\bf 77}, 367 (2017), 1612.03016.

\bibitem{Aaboud:2017svj}
ATLAS, M.~Aaboud {\em et~al.},
\newblock Eur. Phys. J. C {\bf 78}, 110 (2018), 1701.07240,
\newblock [Erratum: Eur.Phys.J.C 78, 898 (2018)].

\bibitem{Aaboud:2017ffb}
ATLAS, M.~Aaboud {\em et~al.},
\newblock JHEP {\bf 12}, 059 (2017), 1710.05167.

\bibitem{Sirunyan:2018swq}
CMS, A.~M. Sirunyan {\em et~al.},
\newblock Eur. Phys. J. C {\bf 78}, 701 (2018), 1806.00863.

\bibitem{Aad:2019rou}
ATLAS, G.~Aad {\em et~al.},
\newblock Eur. Phys. J. C {\bf 79}, 760 (2019), 1904.05631.

\bibitem{Sirunyan:2019bzr}
CMS, A.~M. Sirunyan {\em et~al.},
\newblock JHEP {\bf 12}, 061 (2019), 1909.04133.

\bibitem{Azzi:2019yne}
P.~Azzi {\em et~al.},
\newblock CERN Yellow Rep. Monogr. {\bf 7}, 1 (2019), 1902.04070.

\bibitem{Martin:2004dh}
A.~D. Martin, R.~G. Roberts, W.~J. Stirling, and R.~S. Thorne,
\newblock Eur. Phys. J. {\bf C39}, 155 (2005), hep-ph/0411040.

\bibitem{Schmidt:2015zda}
C.~Schmidt, J.~Pumplin, D.~Stump, and C.~P. Yuan,
\newblock Phys. Rev. {\bf D93}, 114015 (2016), 1509.02905.

\bibitem{Ball:2013hta}
NNPDF, R.~D. Ball {\em et~al.},
\newblock Nucl. Phys. {\bf B877}, 290 (2013), 1308.0598.

\bibitem{Giuli:2017oii}
xFitter Developers' Team, F.~Giuli {\em et~al.},
\newblock Eur. Phys. J. {\bf C77}, 400 (2017), 1701.08553.

\bibitem{Harland-Lang:2016apc}
L.~A. Harland-Lang, V.~A. Khoze, and M.~G. Ryskin,
\newblock Eur. Phys. J. {\bf C76}, 255 (2016), 1601.03772.

\bibitem{Harland-Lang:2016kog}
L.~A. Harland-Lang, V.~A. Khoze, and M.~G. Ryskin,
\newblock Phys. Rev. {\bf D94}, 074008 (2016), 1607.04635.

\bibitem{Manohar:2016nzj}
A.~Manohar, P.~Nason, G.~P. Salam, and G.~Zanderighi,
\newblock Phys. Rev. Lett. {\bf 117}, 242002 (2016), 1607.04266.

\bibitem{Manohar:2017eqh}
A.~V. Manohar, P.~Nason, G.~P. Salam, and G.~Zanderighi,
\newblock JHEP {\bf 12}, 046 (2017), 1708.01256.

\bibitem{Budnev:1974de}
V.~M. Budnev, I.~F. Ginzburg, G.~V. Meledin, and V.~G. Serbo,
\newblock Phys.Rept. {\bf 15}, 181 (1975).

\bibitem{Anlauf:1991wr}
H.~Anlauf, H.~D. Dahmen, P.~Manakos, T.~Mannel, and T.~Ohl,
\newblock Comput. Phys. Commun. {\bf 70}, 97 (1992).

\bibitem{Blumlein:1993ef}
J.~Blumlein, G.~Levman, and H.~Spiesberger,
\newblock J. Phys. {\bf G19}, 1695 (1993).

\bibitem{Mukherjee:2003yh}
A.~Mukherjee and C.~Pisano,
\newblock Eur. Phys. J. {\bf C30}, 477 (2003), hep-ph/0306275.

\bibitem{Luszczak:2015aoa}
M.~Łuszczak, W.~Schäfer, and A.~Szczurek,
\newblock Phys. Rev. {\bf D93}, 074018 (2016), 1510.00294.

\bibitem{Harland-Lang:2019pla}
L.~A. Harland-Lang, A.~D. Martin, R.~Nathvani, and R.~S. Thorne,
\newblock Eur. Phys. J. {\bf C79}, 811 (2019), 1907.02750.

\bibitem{Bertone:2017bme}
NNPDF, V.~Bertone, S.~Carrazza, N.~P. Hartland, and J.~Rojo,
\newblock SciPost Phys. {\bf 5}, 008 (2018), 1712.07053.

\bibitem{Harland-Lang:2019eai}
L.~Harland-Lang,
\newblock JHEP {\bf 03}, 128 (2020), 1910.10178.

\bibitem{Han:1992hr}
T.~Han, G.~Valencia, and S.~Willenbrock,
\newblock Phys. Rev. Lett. {\bf 69}, 3274 (1992), hep-ph/9206246.

\bibitem{Figy:2003nv}
T.~Figy, C.~Oleari, and D.~Zeppenfeld,
\newblock Phys. Rev. {\bf D68}, 073005 (2003), hep-ph/0306109.

\bibitem{Bolzoni:2010xr}
P.~Bolzoni, F.~Maltoni, S.-O. Moch, and M.~Zaro,
\newblock Phys. Rev. Lett. {\bf 105}, 011801 (2010), 1003.4451.

\bibitem{Cacciari:2015jma}
M.~Cacciari, F.~A. Dreyer, A.~Karlberg, G.~P. Salam, and G.~Zanderighi,
\newblock Phys. Rev. Lett. {\bf 115}, 082002 (2015), 1506.02660,
\newblock [Erratum: Phys. Rev. Lett.120,no.13,139901(2018)].

\bibitem{Dreyer:2016oyx}
F.~A. Dreyer and A.~Karlberg,
\newblock Phys. Rev. Lett. {\bf 117}, 072001 (2016), 1606.00840.

\bibitem{Cruz-Martinez:2018rod}
J.~Cruz-Martinez, T.~Gehrmann, E.~W.~N. Glover, and A.~Huss,
\newblock Phys. Lett. {\bf B781}, 672 (2018), 1802.02445.

\bibitem{Liu:2019tuy}
T.~Liu, K.~Melnikov, and A.~A. Penin,
\newblock Phys. Rev. Lett. {\bf 123}, 122002 (2019), 1906.10899.

\bibitem{Dreyer:2020urf}
F.~A. Dreyer, A.~Karlberg, and L.~Tancredi,
\newblock JHEP {\bf 10}, 131 (2020), 2005.11334.

\bibitem{Buonocore:2020nai}
L.~Buonocore, P.~Nason, F.~Tramontano, and G.~Zanderighi,
\newblock JHEP {\bf 08}, 019 (2020), 2005.06477.

\bibitem{Bertone:2015lqa}
V.~Bertone, S.~Carrazza, D.~Pagani, and M.~Zaro,
\newblock JHEP {\bf 11}, 194 (2015), 1508.07002.

\bibitem{Harland-Lang:2020veo}
L.~Harland-Lang, M.~Tasevsky, V.~Khoze, and M.~Ryskin,
\newblock Eur. Phys. J. C {\bf 80}, 925 (2020), 2007.12704.

\bibitem{Bernauer:2013tpr}
A1, J.~C. Bernauer {\em et~al.},
\newblock Phys. Rev. {\bf C90}, 015206 (2014), 1307.6227.

\bibitem{Osipenko:2003bu}
CLAS, M.~Osipenko {\em et~al.},
\newblock Phys. Rev. {\bf D67}, 092001 (2003), hep-ph/0301204.

\bibitem{Airapetian:2011nu}
HERMES, A.~Airapetian {\em et~al.},
\newblock JHEP {\bf 05}, 126 (2011), 1103.5704.

\bibitem{Christy:2007ve}
M.~E. Christy and P.~E. Bosted,
\newblock Phys. Rev. {\bf C81}, 055213 (2010), 0712.3731.

\bibitem{Bertone:2013vaa}
V.~Bertone, S.~Carrazza, and J.~Rojo,
\newblock Comput. Phys. Commun. {\bf 185}, 1647 (2014), 1310.1394.

\bibitem{Harland-Lang:2019zur}
L.~Harland-Lang, J.~Jaeckel, and M.~Spannowsky,
\newblock Phys. Lett. {\bf B793}, 281 (2019), 1902.04878.

\bibitem{Zyla:2020zbs}
Particle Data Group, P.~Zyla {\em et~al.},
\newblock PTEP {\bf 2020}, 083C01 (2020).

\bibitem{Xie:2021equ}
CTEQ-TEA, K.~Xie {\em et~al.},
\newblock (2021), 2106.10299.

\bibitem{Duhr:2020seh}
C.~Duhr, F.~Dulat, and B.~Mistlberger,
\newblock Phys. Rev. Lett. {\bf 125}, 172001 (2020), 2001.07717.

\bibitem{Duhr:2020sdp}
C.~Duhr, F.~Dulat, and B.~Mistlberger,
\newblock JHEP {\bf 11}, 143 (2020), 2007.13313.

\bibitem{Chen:2021vtu}
X.~Chen {\em et~al.},
\newblock (2021), 2107.09085.

\bibitem{Alwall:2014hca}
J.~Alwall {\em et~al.},
\newblock JHEP {\bf 07}, 079 (2014), 1405.0301.

\bibitem{Frederix:2018nkq}
R.~Frederix {\em et~al.},
\newblock JHEP {\bf 07}, 185 (2018), 1804.10017.

\bibitem{Aad:2015auj}
ATLAS, G.~Aad {\em et~al.},
\newblock Eur. Phys. J. {\bf C76}, 291 (2016), 1512.02192.

\bibitem{Bizon:2018foh}
W.~Bizoń {\em et~al.},
\newblock JHEP {\bf 12}, 132 (2018), 1805.05916.

\bibitem{Kallweit:2015fta}
S.~Kallweit, J.~M. Lindert, S.~Pozzorini, M.~Sch\"onherr, and P.~Maierh\"ofer,
\newblock {NLO QCD+EW automation and precise predictions for V+multijet
  production},
\newblock in {\em {50th Rencontres de Moriond on QCD and High Energy
  Interactions}}, 2015, 1505.05704.

\bibitem{Bozzi:2011ww}
G.~Bozzi, J.~Rojo, and A.~Vicini,
\newblock Phys. Rev. D {\bf 83}, 113008 (2011), 1104.2056.

\bibitem{Bozzi:2015hha}
G.~Bozzi, L.~Citelli, and A.~Vicini,
\newblock Phys. Rev. D {\bf 91}, 113005 (2015), 1501.05587.

\bibitem{Bagnaschi:2019mzi}
E.~Bagnaschi and A.~Vicini,
\newblock (2019), 1910.04726.

\bibitem{Hussein:2019kqx}
M.~Hussein, J.~Isaacson, and J.~Huston,
\newblock J. Phys. G {\bf 46}, 095002 (2019), 1905.00110.

\bibitem{Farry:2019rfg}
S.~Farry, O.~Lupton, M.~Pili, and M.~Vesterinen,
\newblock Eur. Phys. J. C {\bf 79}, 497 (2019), 1902.04323.

\bibitem{CarloniCalame:2016ouw}
C.~M. Carloni~Calame {\em et~al.},
\newblock Phys. Rev. D {\bf 96}, 093005 (2017), 1612.02841.

\bibitem{Sjostrand:2007gs}
T.~Sjostrand, S.~Mrenna, and P.~Z. Skands,
\newblock Comput. Phys. Commun. {\bf 178}, 852 (2008), 0710.3820.

\bibitem{Aad:2014xaa}
ATLAS, G.~Aad {\em et~al.},
\newblock JHEP {\bf 09}, 145 (2014), 1406.3660.

\bibitem{Campbell:2002tg}
J.~M. Campbell and R.~Ellis,
\newblock Phys. Rev. D {\bf 65}, 113007 (2002), hep-ph/0202176.

\bibitem{Campbell:2004ch}
J.~M. Campbell, R.~Ellis, and F.~Tramontano,
\newblock Phys. Rev. D {\bf 70}, 094012 (2004), hep-ph/0408158.

\bibitem{PDG2019}
Particle Data Group, M.~Tanabashi {\em et~al.},
\newblock Phys. Rev. D {\bf 98}, 030001 (2018).

\bibitem{Bacchetta:2018lna}
A.~Bacchetta, G.~Bozzi, M.~Radici, M.~Ritzmann, and A.~Signori,
\newblock Phys. Lett. B {\bf 788}, 542 (2019), 1807.02101.

\bibitem{Collins:1977iv}
J.~C. Collins and D.~E. Soper,
\newblock Phys. Rev. D {\bf 16}, 2219 (1977).

\bibitem{ATLAS:2014kca}
ATLAS, G.~Aad {\em et~al.},
\newblock JHEP {\bf 03}, 041 (2015), 1412.0237.

\bibitem{Aaboud:2017qph}
ATLAS, M.~Aaboud {\em et~al.},
\newblock Eur. Phys. J. C {\bf 78}, 199 (2018), 1710.09748.

\bibitem{Sirunyan:2018xiv}
CMS, A.~M. Sirunyan {\em et~al.},
\newblock JHEP {\bf 01}, 122 (2019), 1806.10905.

\bibitem{Sirunyan:2020ztc}
CMS, A.~M. Sirunyan {\em et~al.},
\newblock Eur. Phys. J. C {\bf 80}, 752 (2020), 2001.10086.

\bibitem{Buckley:2014ana}
A.~Buckley {\em et~al.},
\newblock Eur. Phys. J. C {\bf 75}, 132 (2015), 1412.7420.

\bibitem{Bertone:2018dse}
V.~Bertone, R.~Gauld, and J.~Rojo,
\newblock JHEP {\bf 01}, 217 (2019), 1808.02034.

\end{thebibliography}
\bibliographystyle{h-physrev}

\end{document}